\documentclass[10pt]{article}

\usepackage{amsmath}
\usepackage{amssymb}
\pdfoutput=1
\usepackage{graphicx,color,subfigure}
\usepackage{multirow,rotating}

\usepackage{cite}
\usepackage{color}

\usepackage{placeins}

\usepackage[labelfont=bf,labelsep=period,justification=raggedright]{caption}

\makeatletter
\renewcommand{\@biblabel}[1]{\quad#1.}
\makeatother

\date{}

\begin{document}

\begin{flushleft}
{\Large
\textbf{Integrative modeling of sprout formation in angiogenesis: coupling the VEGFA-Notch signaling in a dynamic stalk-tip cell selection}
}
\\
Sotiris A.Prokopiou$^{1\ast}$,
Markus R.Owen$^{2}$,
Helen M.Byrne$^{3}$,
Safiyyah Ziyad$^{4}$,
Courtney Domigan$^{4}$,
M.Luisa Iruela-Arispe$^{4}$,
Yi Jiang$^{5}\dagger$
\\
\bf{1} Inria team Dracula, Inria center Grenoble Rhone-Alpes, 66 Boulevard Niels Bohr F-69603 Villeurbanne, France
\\
\bf{2} Centre for Mathematical Medicine and Biology, School of Mathematical Sciences, University of Nottingham, Nottingham NG7 2RD, UK
\\
\bf{3} Oxford Centre for Collaborative Applied Mathematics, University of Oxford, Oxford OX1 3LB, UK and Department of Computer Science, University of Oxford, Oxford OX1 3QD, UK
\\
\bf{4} Department of Molecular, Cell, and Developmental Biology, University of California Los Angeles, Los Angeles, CA 90095, USA
\\
\bf{5} Department of Mathematics and Statistics, Georgia State University, Atlanta, Georgia 30303, USA
\\
 E-mail: $\ast$sotirisprokopiou1@gmail.com; $\dagger$yjiang12@gsu.edu; markus.owen@nottingham.ac.uk.
\end{flushleft}

\section*{Abstract}
During angiogenesis, new blood vessels headed by a migrating endothelial tip cell sprout from pre-existing ones. This process is known to be regulated by two signaling pathways concurrently, vascular endothelial growth factor A (VEGFA) and Notch-Delta.
Extracellular VEGFA activates the intracellular Notch-Delta pathway in nearby endothelial cells which results in endothelial (stalk, tip) differentiation. Retinal astrocytes appear to play a crucial role in polarizing new sprouts by secreting VEGFA.
\emph{In vivo} retinal angiogenesis experiments in neonatal mouse generated quantitative data on daily cell counts and morphological data of vascular network expanding over fibronectin-rich matrix.
Based on this set of data and other existing ones, we developed a cell-based, multiscale mathematical model using the cellular Potts model framework to investigate the sprout evolution by integrating the VEGFA and Notch-Delta signaling pathways.
The model incorporates three levels of description:
1) intracellular: each endothelial cell incorporates the VEGFA-activated Notch-Delta signaling pathway,
2) intercellular: tip and stalk endothelial phenotypes along the sprout are dynamically (depending on cell growth, cell-cell adhesion) interchangeable based on their Delta level, and
3) extracellular: the astrocyte-derived VEGFA acts as a chemoattractant and, together with the extracellular matrix (ECM) network, guides cell migration leading to sprout evolution and morphology.
Starting with a single astrocyte embedded in a fibronectin-rich matrix, we use the model to assess different scenarios regarding VEGFA levels and its interaction with matrix proteins. Simulation results suggest that astrocyte-derived VEGFA gradients along with heterogeneous ECM reproduces sprouting morphology, and the extension speed is in agreement with experimental data in 7 days postnatal mouse retina.
Results also reproduce empirical observations in sprouting angiogenesis, including anastomosis, dynamic tip cell competition, and sprout regression as a result of Notch blockade.
\section*{Author Summary}
Growth of new blood vessels from existing ones, called angiogenesis, is important in normal development, wound healing, and many pathological processes including cancer and macular degeneration. Our goal was to understand the evolution of sprout formation.
Peripheral retinal astrocytes secrete VEGFA which is coupled with the intracellular Delta-Notch signaling pathway of the endothelial cells in a nearby blood vessel.
We developed a multiscale cellular Potts model of sprouting angiogenesis, focusing on the VEGFA-Delta-Notch signaling pathway, cell differentiation (between stalk and tip endothelial phenotypes), growth, migration, cell-cell, and cell-matrix adhesion.
Simulation results suggest that VEGFA gradients and heterogeneous extracellular matrix are key for sprout formation.

\section*{Introduction}
\subsection*{Biological background}
Angiogenesis, the formation of blood vessels from a pre-existing vasculature, is a process whereby capillary sprouts form in response to externally supplied angiogenic stimuli. The new sprouts provide tissues and organs with oxygen and nutrients, and remove metabolic waste. Angiogenesis takes place in physiological situations, such as embryonic development, wound healing and reproduction \cite{Carmeliet2005}. The healthy body controls angiogenesis by balancing pro- and anti-angiogenic factors \cite{CarmelietJain2000}. This balance, though, is sometimes disrupted and angiogenesis also appears in many pathologies, like diabetes \cite{Martin2003}, rheumatoid arthritis \cite{Koch2003}, cardiovascular ischemic complications \cite{Cao2005}, proliferative retinopathy \cite{Retinopathy1990}, and cancer \cite{Folkman1975}.

Sprouting angiogenesis typically starts from hypoxic tissues or cells (e.g. retinal astrocytes \cite{Scott2010}) upregulating their production of pro-angiogenic factors. The primary type is the vascular endothelial growth factor A (VEGFA) \cite{Folkman1971}. These angiogenic factors diffuse and bind to the endothelial cell (EC) receptors of nearby blood vessels.
Subsequently, the extracellular matrix (ECM) and basement membrane, surrounding the ECs, are degraded locally by activated proteases (e.g. matrix metalloproteinases, MMPs) produced by ECs.

\begin{figure}[h!]
\begin{center}
  \includegraphics[width=3.5in]{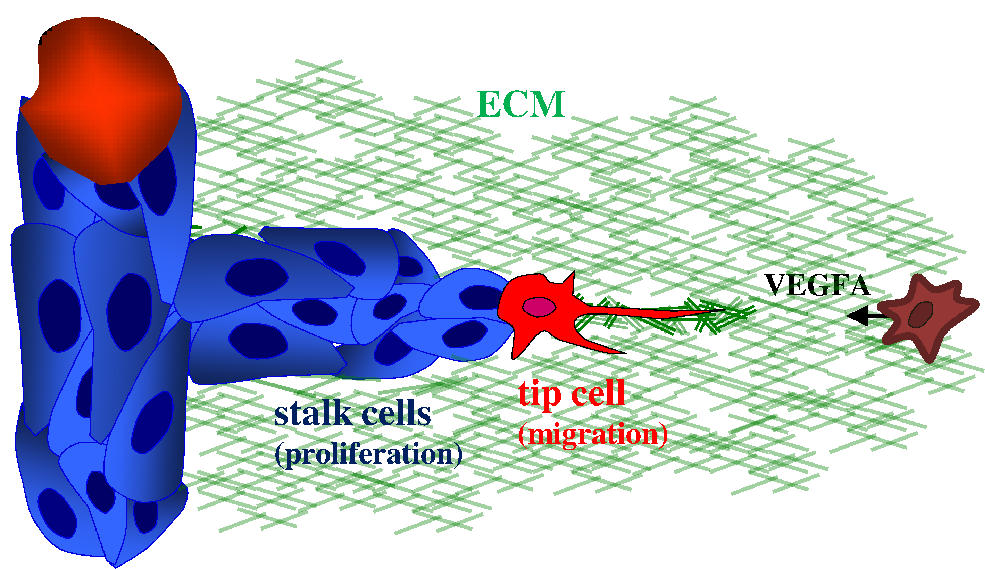} 
\end{center}
\caption{\textbf{Schematic of processes involved in a growing vascular sprout.} A growing sprout is mainly composed of tip (red) and stalk (blue) cells, where each endothelial phenotype responds differently (e.g. proliferation, migration) to a VEGFA source (astrocyte (brown)). ECM = extracellular matrix.}\label{fig:schematic_sprout_morphology}
\end{figure}

Several experimental studies (e.g. in mouse retina \cite{Gariano2005}) have revealed the heterogeneity of a growing vascular sprout. It is composed of a tip cell at the leading front, followed by stalk cells. The latter cells comprise the body of the sprout and are responsible for the elongation of the new vascular branch (Figure \ref{fig:schematic_sprout_morphology}).
The tip and stalk EC phenotypes display different gene expression profiles, suggesting that their specification is determined genetically \cite{delToro2010}. However, a single unique gene or protein that can be used reliably and unambiguously as a molecular marker for those phenotypes has not been identified.
A key pathway, though, regulating their specification is the Notch-Delta signaling pathway which is described next. This finding concurrently emerged from several groups using three distinct experimental models of angiogenesis including solid tumor in mice \cite{Ridgway2006,Troise2006}, postnatal mouse retinas \cite{Hellstrom2007,Suchting2007,Lobov2007}, and zebrafish embryos \cite{Siekmann2007,Leslie2007}. Our model is mainly motivated from retina angiogenesis data on the post-natal day 7, where vascular sprouts are already developed \cite{Dorrell2002}.

Notch-Delta signaling is a fundamental signaling pathway between neighboring cells.
It has a role in processes that include lateral-inhibition \cite{Shaya2011}, synchronising cells during somitogenesis \cite{Dequeant2008}, asymmetric cell division \cite{Roegiers2004,Bardin2004}, and neuronal plasticity \cite{Lieber2011,Alberi2011}.
Notch-Delta is mediated by the interaction between Notch receptors and Delta/Serrate/LAG-2 (DSL) ligands \cite{Bray2006}. Upon interaction between a Notch receptor in one cell and a DSL ligand in a neighboring cell, the Notch intracellular domain is cleaved, translocates to the nucleus and co-activates downstream transcriptional targets.
Notch signaling also plays an important role in fine-grained patterning processes such as the formation of checkerboard-like differentiation patterns and sharp boundaries between developing tissues \cite{Shaya2011}.

In this study, we are mostly interested in the lateral-inhibition mechanism, which is responsible for
generating an alternating pattern between stalk and tip EC phenotypes (`salt-pepper' pattern) \cite{Hellstrom2007}.
Sprout formation requires the coordinated behavior of all EC phenotypes. The head tip cell located at the leading edge of the growing vessel senses the environment for angiogenic factors through the presence of filopodia. In contrast, the stalk cells situated behind the tip cells are highly proliferative cells and allow the vessel to elongate towards angiogenic stimuli.
Endothelial tip cells are stimulated by an extracellular gradient of VEGFA (and/or VEGFC \cite{Geudens2011}). ECs express three different VEGF receptors (VEGFR). The activity of VEGFR2 (Flk-1) regulates most of the EC responses to VEGFA, including induction of tip cell filopodia and EC migration, proliferation, survival, and vascular permeability.

Recent \emph{in vivo} and \emph{in vitro} studies from Gerhardt \emph{et al.} \cite{Jakobsson2010} on retina sprout angiogenesis revealed that ECs dynamically compete with each other for the tip cell position. In particular, it was shown that VEGFR2 levels between two cells affect which of them will become a tip cell, in a competitive manner. The authors suggested that the balance between VEGFR2 and VEGFR1 (Flt-1) expression in individual ECs affects their potential to become tip cells during sprouting angiogenesis. That is, cells with higher VEGFR2 levels stand a better chance to take and maintain the leading position.
The head tip cell is dynamically challenged and replaced by migrating cells from the stalk region. This dynamic competition between ECs for a tip or stalk phenotype depends on the integrated VEGFA-Delta-Notch signaling pathway, the main focus of the current work.
One of the main findings of recent years has been the identification of the Notch-Delta pathway as the instructive regulator of tip versus stalk cell fate \cite{Arispe2009}.
Both the receptor (e.g. Notch1) and ligand (Delta-like ligand 4; Dll4) are cell bound and thus act only through cell-cell contact. VEGFA-VEGFR2 signaling was shown to upregulate expression of Dll4 in tip cells \cite{Lobov2007}, allowing Dll4 to activate Notch1 in the adjacent stalk cells, causing suppression of the tip cell phenotype. Hence, ECs exposed to the highest VEGFA concentration are most likely to become tip cells. 

Strong evidence has shown that astrocytes are a major producer of VEGF \cite{Scott2010,Astrocytes2005}. Moreover the pronounced expression of VEGF in peripheral astrocytes has been linked to sprouting angiogenesis at the leading edge of the expanding vascular plexus \cite{Scott2010}. Macrophages are also a major source of diffusible VEGF, but the fact that they are located behind the leading edge is linked to sprout fusion \cite{BaerArispe2013}.
Another important ingredient with many roles in angiogenesis is the ECM, a mesh-like network of proteins. It is essential for EC migration, proliferation and survival, since it provides structural support and chemical cues for cell adhesion and motility \cite{Arispe1991}. ECM components like collagen I and fibrin are capable of supporting chemotactic migration \cite{Retinopathy1990}. The density and spatial distribution of ECM proteins such as fibronectin and collagen can affect the speed and direction of cell migration \cite{DiMilla1993}. Furthermore, ECs are able to secrete and degrade ECM components.
ECs activated by VEGFA first degrade the basement membrane of the parent vessel and then migrate into the ECM towards the VEGFA source (e.g. astrocyte).
The local degradation (e.g. via MMPs) and deposition of matrix proteins by ECs and the heterogeneity of the ECM can all create local gradients of ECM components which can drive EC migration, a process called haptotaxis.


\FloatBarrier
\subsection*{Computational background}
Mathematical modeling in angiogenesis is a useful tool for understanding the interplay between the factors that affect it and for the design of experiments of a predictive nature.

Over the past two decades, a plethora of mathematical and computational models have been developed to study different aspects of angiogenesis. In general, these models could be divided into two main groups: continuous and discrete models.
Briefly, continuous models describe ECs as densities using partial differential equations \cite{Hogea2006,Zheng2012}. However, because the vessels of vascular networks often consist of only a few cells, explicitly considering individual cells is essential. Discrete models describe ECs as entities where their behaviors can be explicitly represented \cite{Chaplain1995,Lauffenburger2010,McDougallChaplain2012}. However, in these models the detailed investigation of cell-level properties, such as cell shape and cell adhesion, are mathematically difficult, if not impossible to consider.
Therefore, we choose to focus on cell-based models, which is a subgroup of discrete models with the advantage of representing cells as individual entities with a particular shape.
For a comprehensive review of mathematical and computational models in angiogenesis see \cite{Peirce2008}, and references therein.

In Table \ref{tab:Models} we summarize a number of mathematical (mainly cell-based) models for studying aspects of angiogenesis which have delivered useful insights for angiogenic sprouting.
In particular, we refer to whether a model has assessed the role of mitosis on sprouting, whether ECM has been considered and if it was simply incorporated as a uniform field or as a non-uniform representation of ECM fibers, the VEGF profile (static gradients or secretion from a source), and also, whether the stalk-tip cell selection decision has been made upon the position of a cell in a sprout (e.g. a cell is simply defined as a tip if it is located at the front of a sprout) or whether the Notch-Delta signaling has been explicitly considered for the EC fate.

Finally, our computational model, discussed next, can be distinguished (from the models presented in Table \ref{tab:Models}) by its integrative approach, meaning that experimental data are used for the validation of the model.

\FloatBarrier
\begin{table}[h]
\caption{Main mechanisms incorporated in sprouting angiogenesis models} 
\centering 
\begin{tabular}{c c c c c c c c} 
\hline\hline 
{\small Authors} & {\small Type of model} & {\small ECM} & {\small Mitosis} & {\small VEGF} & {\small stalk/tip}  & {\small Delta/} & {\small D/N}  \\ [0.25ex]
                &               &     &         &      & {\small phenotype}  & {\small Notch}  & {\small blockade} \\ [0.5ex]
\hline\hline 
{\small Bauer}   & {\small 2-D Cellular}  & {\small non-uniform}  & $\checkmark$  & $\checkmark$  & $\text{\sffamily X}$  & $\text{\sffamily X}$  & $\text{\sffamily X}$  \\[0.001cm]
{\small$\emph{et al.}$} \cite{Bauer2007} & {\small Potts Model}  & {\small(ECM fibers)}  &                        & {\small secreted from} &   &   &   \\[0.001cm]
                                       &              &               &                        & {\small source}        &   &   &   \\[1ex]
\hline 
{\small Milde}   & {\small 3-D particle model}  & {\small non-uniform} & $\checkmark$ & $\checkmark$ & $\checkmark$  & $\text{\sffamily X}$  & $\text{\sffamily X}$  \\[0.25ex]
{\small$\emph{et al.}$} \cite{Koumoutsakos2008} & {\small(stalk cell density,}     & {\small(ECM fibers,}     &    & {\small bound-} $\&$  & {\small(one leading}  &   &   \\[0.25ex]
&  {\small particle}   &  {\small ECM proteolysis)}    &    &  {\small soluble-} & {\small tip cell)}  &   &   \\[0.25ex]
& {\small representation}  &  &    & {\small VEGF isoforms}   &      &   &   \\[0.25ex]
& {\small of the tip cell)}  &  &    &    &      &   &   \\[1ex]
\hline 
{\small Bentley}   & {\small 3-D (multi-agent)}  & $\text{\sffamily X}$  & $\text{\sffamily X}$ & $\checkmark$ & $\checkmark$ & $\checkmark$ & $\checkmark$  \\[0.25ex]
{\small$\emph{et al.}$} \cite{Bentley2008}  & {\small lattice model}      &    &  & {\small static } & {\small(`salt-pepper'} &  &   \\[0.25ex]
&                             &    &  & {\small gradients} & {\small pattern)} &  &   \\[1ex]
\hline 
{\small Bentley}      & {\small 3-D (multi-agent)}  & $\text{\sffamily X}$  & $\text{\sffamily X}$  & $\checkmark$ & $\checkmark$ & $\checkmark$ & $\text{\sffamily X}$ \\[0.25ex]
{\small$\emph{et al.}$} \cite{Bentley2009} & {\small lattice model}      &  &  & {\small static} & {\small(`salt-pepper'} &  &   \\[0.25ex]
&                             &  &  & {\small gradients} & {\small pattern)} &  &   \\[1ex]
\hline 
{\small Qutub}     & {\small 3-D lattice}  & {\small uniform}                  & $\checkmark$ & $\checkmark$ & $\checkmark$  & $\checkmark$ & $\checkmark$ \\[0.5ex]
{\small$\emph{et al.}$} \cite{Qutub2009}  &  {\small model}   & {\small(collagen}        &   & {\small static} & {\small(one leading} & & \\[0.25ex]
&              & {\small concentration)}  &   & {\small gradients} & {\small tip cell)} & & \\[1ex]
\hline 
{\small Das}  & {\small 3-D lattice}  & {\small uniform}                  & $\checkmark$ & $\checkmark$ & $\text{\sffamily X}$  & $\text{\sffamily X}$  & $\text{\sffamily X}$  \\[0.25ex]
{\small$\emph{et al.}$} \cite{Lauffenburger2010}  &  {\small model}     & {\small(collagen} &              & {\small bound-} $\&$ &   &   &   \\[0.25ex]
&     &  {\small concentration,}  &    &  {\small soluble-} &   &   &   \\[0.25ex]
&     & {\small ECM proteolysis)}   &    & {\small VEGF isoforms} &   &   &   \\[0.25ex]
\hline 
{\small McDougall}  & {\small 2-D hybrid}  & {\small uniform} & $\checkmark$ & $\checkmark$ & $\checkmark$  & $\text{\sffamily X}$  & $\text{\sffamily X}$  \\[0.25ex]
{\small$\emph{et al.}$} \cite{McDougallChaplain2012}  & {\small model} &  {\small(fibronectin} &    & {\small secreted by} & {\small(one leading} & & \\[0.25ex]
  &  &  {\small concentration} & & {\small astrocytes} & {\small tip cell)} & & \\[0.25ex]
 &               &   {\small secreted by}    & &  &     & & \\[0.25ex]
 &   &   {\small  astrocytes,}    &        &  &    & & \\[0.25ex]
 &                    &   {\small ECM proteolysis)}          &  &            &              & & \\[0.25ex]
\hline 
{\small Daub and}   & {\small 2-D Cellular}  & {\small uniform}  & $\checkmark$  & $\checkmark$  & $\checkmark$  & $\text{\sffamily X}$  & $\text{\sffamily X}$  \\[0.001cm]
{\small Merks} \cite{Daub2013} & {\small Potts Model}  & {\small(collagen}         &   & {\small static} & {\small (one leading}   &   &   \\[0.001cm]
&                       & {\small concentration,}   &   & {\small gradients} & {\small tip cell)}  &   &   \\[0.001cm]
                                       & & {\small ECM proteolysis)} &   &  &   &   &   \\[1ex]
\hline\hline
\end{tabular}
\label{tab:Models}
\end{table}

\FloatBarrier
\section*{Methods}
The aim of the current work is to expand our understanding of sprout evolution through the VEGFA-Dll4-Notch1 signaling pathway along with the dynamic competition between stalk and tip cells for the tip cell position.
We incorporate all of the mechanisms from Table \ref{tab:Models}, and in doing so, we improve upon other sprouting angiogenesis models.
In this section we describe the three different levels of our cell-based, multiscale model.

\FloatBarrier
\subsection*{Cellular level: the Cellular Potts Model}
The cellular model is based on the Cellular Potts Model (also known as the Glazier-Graner-Hogeweg model or GGH) \cite{Glazier1993}. The cell interactions are characterized through a total energy (or Hamiltonian, $H$)
\begin{equation}\label{eq:Hamiltonian}
H =\sum_{(\vec{x},\vec{x'})} J_{\tau(\sigma(\vec{x})),\tau(\sigma(\vec{x'}))}\left(1- \delta_{\sigma(\vec{x}),\sigma(\vec{x'})}\right) + \lambda_{area} \sum_{\sigma>0} \left(a_{\sigma}-A_{\sigma}\right)^{2} + H',
\end{equation}
where $J$ are the contact energies between cells, $a_{\sigma}$ the cell area and $A_{\sigma}$ the target cell area. In addition, $\tau(\sigma)$ represents the type of the cell occupying a grid space $\sigma$, which in our model can be either the astrocyte or one of the two EC phenotypes (stalk and tip cells). The Kronecker delta function is $\delta_{x,y}={1, \textrm{if}\:\: x=y; 0, \textrm{if}\:\: x \neq y}$, and the term $(1- \delta_{\sigma(\vec{x}),\sigma(\vec{x'})})$ ensures that adhesive energy only accrues at cell surfaces (not inside the cells). $H'$ can be any constraint on the cell behavior (e.g. chemotaxis \cite{Savill1997}, haptotaxis \cite{Turner2002}, cell elongation \cite{Merks2006}).
To mimic cytoskeletally driven, active surface fluctuations, a lattice site $\vec{x}$ and a neighboring target $\vec{x'}$
are randomly selected.
Then we calculate how the effective energy would change if the initial site displaced the target. If the energy decreases ($\Delta H=H_{new}-H_{old}<0$), the change occurs with probability 1. However, if $H$ increases ($\Delta H>0$), the change will be accepted with Boltzmann probability, $p=\exp(-\Delta H/T)$, where $T$ is the `temperature' (like thermal fluctuations in statistical physics \cite{Beysens2000}) of the system; $T$ influences the likelihood of energetically unfavorable events taking place: the higher $T$, the more out-of-equilibrium the system will be.
Biologically, $T$ represents the amplitude of cell membrane fluctuations (and not active cell movement). On a lattice with $M$ sites, $M$ site copy attempts represent our basic unit of time, one Monte Carlo Step (MCS).

In Table \ref{table:modelmechanisms} we summarize the different behaviors of the stalk and tip cells, which are discussed in detail in the following sections.

\FloatBarrier
\begin{table}[ht]
\caption{Stalk and tip cell behaviors} 
\centering 
\begin{tabular}{l l l l} 
\hline\hline 
    & \textbf{Behavior} & \textbf{Stalk cells} & \textbf{Tip cells} \\ [1ex] 
\hline 
1.  & chemotaxis       & weak                             & strong \\ [1ex]
2.  & growth           & if adjacent to tip cells         & none \\ [1ex]
3.  & mitosis          & if cell area$(a(t))>2a(0)$ and      & none \\ [1ex]
    &                  & cell cycle time $>$17 hrs        &  \\ [1ex]
4.  & elongation       & target length = $a$/2            & target length = $a$/2 \\ [1ex]
    &                  & (if adjacent to tip cells)       &   \\ [1ex]
5.  & switch phenotype & become tip                       & become stalk  \\ [1ex]
    &                  & if Delta$>$threshold($D^*$)      & if Delta$<$threshold($D^*$) \\ [1ex]
\hline 
\end{tabular}
\label{table:modelmechanisms} 
\end{table}

\begin{figure}[h!]
\begin{center}
\includegraphics[width=0.3\textwidth,angle=0]{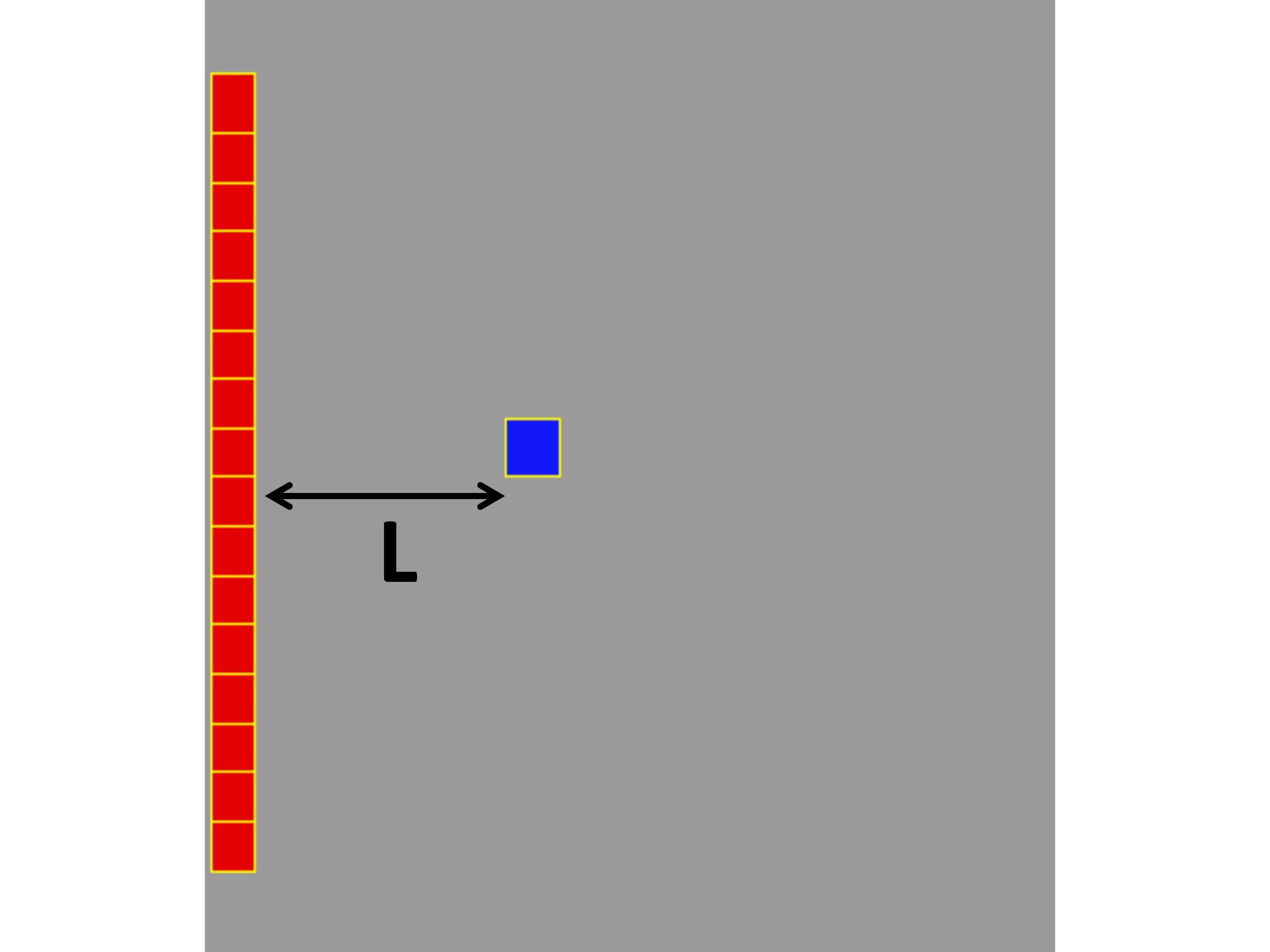}
\end{center}
\caption{\textbf{Model setup.} Model setup on the numerical domain with astrocyte (VEGFA source). Key: stalk cells (red); astrocyte (blue).}\label{fig:setup}
\end{figure}

For the computational results presented in the next section, we use the open-source simulation environment CompuCell3D (CC3D) \cite{Glazier1993}.
We ran our simulation on a 200$\mu m$ x 200$\mu m$ lattice (with no flux boundary conditions) as depicted in Figure \ref{fig:setup} with the parent vessel located along one side and the astrocyte located at distance $L$ from the blood vessel. Although an astrocyte might be found ($\sim$10 $\mu m$) or ($>$20 $\mu m$) away from a blood vessel \cite{McCaslin2011}, we considered $L=60 \mu m$ such that the stalk-tip cell dynamics could be clearly visible and captured along the new sprout formation.

\FloatBarrier

By equating the initial cell surface area in the model to the real cell size (e.g. 5x5 pixels = 10x10 $\mu m^2$ \cite{Wiltbank1994}), we can convert the lattice spacing to microns (1 pixel = 2 $\mu m$; 1 pixel represents square region of length 2 $\mu m$). In addition, we set the time conversion to be 1 MCS = 0.01 hrs, a value
similar to those used in other CPM studies \cite{Bauer2007,Chaplain2000}.

\FloatBarrier
\subsubsection*{Mitosis}
\label{subsec:mitosis}
It is well accepted that in the absence of EC proliferation, angiogenesis is incomplete,
that is, the ECs fail to reach their target \cite{Anderson1998}. Sprouting is possible without EC proliferation only up to a limited extension length.
Proliferation is necessary to sustain sprouting for a longer period and to produce a sprout which is long enough to reach its target \cite{Gerhardt2008}.

After tip cell activation by VEGFA, small sprouts form by aggregation and migration of ECs that are recruited from the parent vessel. The sprout further extends when ECs in the sprout begin to divide \cite{Folkman1987,Ausprunk1977}.
Tip cells rarely proliferate \cite{Gerhardt2003} compared to the strongly proliferative stalk cells, which support sustained elongation of the growing sprout \cite{Gerhardt2003}.

In our cell level model, we postulate that only stalk cells adjacent to tip cells proliferate, because otherwise (if all stalk cells divide) that would lead to a thick/swollen parent vessel. In an attempt to avoid any predefined or probabilistic rules (as in \cite{Qutub2009}), each cell carries a clock defined as
\begin{equation}
\dfrac{d\phi}{dt} =
\left\{
\begin{array}{ll}
a_1 &, \textrm{if cell is stalk adjacent to tip},\\
0   &, \textrm{otherwise},
\end{array}
\right.
\label{eq:clock}
\end{equation}
which progresses only when the cell is a stalk cell that is adjacent to a tip cell, and $a_1$ is chosen to be the time conversion between the real and \emph{in silico} time ($a_1=0.01$ hrs/MCS). In addition, the target area of a cell ($A$) grows with a rate $\mu=\frac{2A-A}{\textrm{cell cycle}}=\frac{25}{1700}=0.0147$ pixels/MCS.
A stalk cell can divide if two conditions apply: 1) its clock reaches the cell cycle duration (17 hrs as evaluated later in the experimental results section, which is very close to the value of 18 hrs used in \cite{Bauer2007,Chaplain2000}), and 2) its cell area doubles. Note that $\phi(0) \in (0,17)$ (at $t=0$ $\phi(0)$ is drawn randomly from a uniform distribution), and when $\phi(t)=17$ it is reset to zero.
%

Cell division involves assigning half of the area to its two daughter cells, and they each inherit the properties (stated in Table \ref{table:modelmechanisms}) of their parent.
Endothelial proliferation might be influenced by VEGF (as modeled in Qutub and Popel \cite{Qutub2009}). For simplicity, however, we assume that the proliferation rate is independent of any external (growth) factors. In other computational models, Milde \emph{et al.} \cite{Koumoutsakos2008} considered proliferation of the tip cell and assumed that a capillary branches when its tip cell divides. On the other hand, Bentley \emph{et al.} \cite{Bentley2009} excluded cell growth and division in their study by focusing exclusively on the early stages of sprouting.

\FloatBarrier
\subsubsection*{Elongation}
To elongate the stalk region, stalk cells must divide, but cell elongation also seems to be important \cite{Merks2006}. Experimental results in \cite{Geudens2011} show that stalk cells in close contact with tip cells appear to elongate. 

In our model, the elongation of tip cells and those stalk cells which are adjacent to tip cells is incorporated by following Merks \emph{et al.} \cite{Merks2006} and including an extra term in the energy equation (\ref{eq:Hamiltonian}) of the form:
\begin{equation}\label{eq:elongation1}
H'_{length} = \lambda_{length} \sum \left( l_{\sigma} - L_{\sigma}(t) \right)^2,
\end{equation}
where $l_{\sigma}$ is the length of cell $\sigma$ along its longest axis, $L_{\sigma(t)}$ its target length, and $\lambda_{length}$ the strength of the length constraint. We do not assign a constant target length as in Merks \emph{et al.} \cite{Merks2006}; instead, we impose a dynamic constraint on the target cell length so that
\begin{equation}\label{eq:elongation2}
L_{\sigma}(t) = a_{\sigma}(t)/2,
\end{equation}
where $a_{\sigma}(t)$ is the current cell area at each MCS. One of the main reasons for implementing tip and stalk cell elongation in the way described above is to prevent cells from elongating before tip cell activation. In addition, by dividing cell area by 2, and by also considering that cells double in size before they divide, it allows the cells to have a length in the range (from $a_{\sigma}(0)/2$ to $a_{\sigma}(0)$).

\FloatBarrier
\subsection*{Extracellular level}
\FloatBarrier
\subsubsection*{ECM implementation and haptotaxis}
\begin{figure}[h!]
\begin{center}
\includegraphics[width=0.3\textwidth,angle=0]{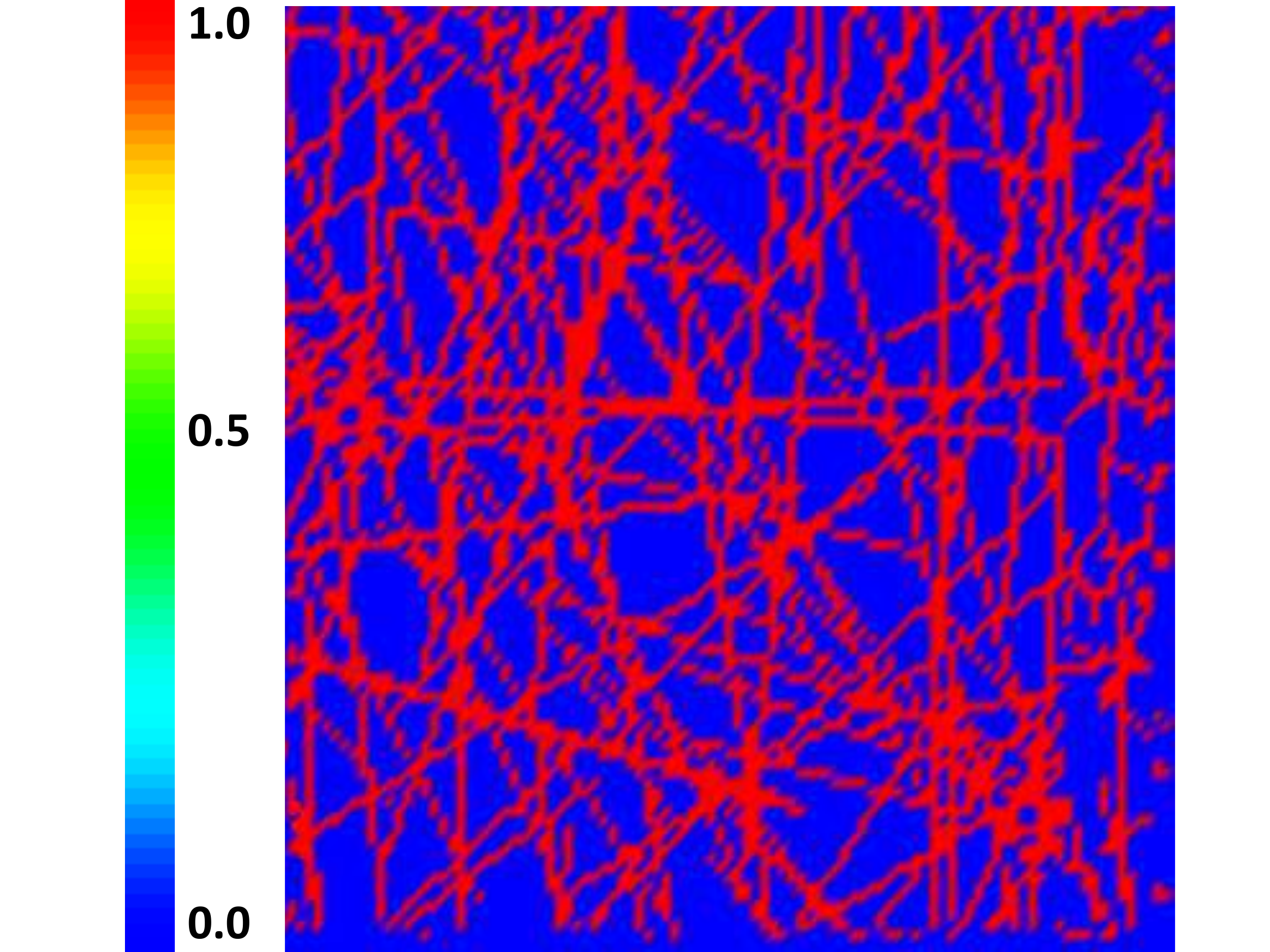}
\end{center}
\caption{\textbf{ECM structure.} Extracellular matrix (ECM) structure as implemented in the CPM framework with the colorbar showing the ECM levels.}\label{fig:setupECM}
\end{figure}

ECM is an important factor with which the cells interact. In CC3D, the substrate (e.g. medium) is normally represented as a fixed cell covering the whole computational domain. In order to model a non-uniform ECM, we distributed ECM fibers randomly in the numerical domain as shown in Figure \ref{fig:setupECM}.
We suppose that ECM fibers are thinner than the cell size (width $=$ 1 pixel), and we assume that the ECM is rigid (fixed in space). That is, cells do not produce or degrade ECM.
Each pixel in the numerical domain occupied by an ECM fiber is given a non zero (=1) value (and zero elsewhere), so that the cells can preferentially adhere to the fibers.

\FloatBarrier

Haptotaxis, the directional migration of cells up ECM gradients, can be incorporated into the model by including the following contribution to the energy term in equation (\ref{eq:Hamiltonian})
\begin{equation}\label{eq:haptokinesis}
H'_{hapt} = \lambda_{ECM}\left( ECM(\vec{x}) - ECM(\vec{x}') \right),
\end{equation}
where, $ECM(\vec{x})$ is the ECM concentration at site $\vec{x}$, and $\lambda_{ECM}$ is the strength of the preferential attachment to ECM.

Experimental studies showed that tip cells must first break through the basement membrane of the parent blood vessel, and afterwards, proteolytically degrade and invade into the ECM in order to form a new capillary \cite{Li2005}. In addition, aligned fibers in the ECM can guide cell migration, and these guiding structures can in turn be remodelled by EC tip cells \cite{Li2005}. Differences in ECM rigidity or stiffness can also direct migration.
However, remodeling of ECM by ECs, stiffness of ECM, and ECM degradation are not considered in our model, since we don't expect stiffness to vary much across the retina, and we don't expect the others to contribute much to the overall process.

The effect of non-uniform ECM on the sprout formation has been modelled previously by Bauer \emph{et al.} \cite{Bauer2007}, and Milde \emph{et al.} \cite{Koumoutsakos2008}. Briefly, in \cite{Bauer2007}, the authors assumed that branch splitting resulted from ECM inhomogeneities, and in \cite{Koumoutsakos2008}, the authors showed that the structure and density of the ECM has a direct effect on the morphology, expansion speed and number of branches.

\FloatBarrier
\subsubsection*{VEGFA: concentration and local gradients}
VEGF is one of the main growth factors involved in angiogenesis. Experimental studies have demonstrated that the absolute VEGF concentration and the VEGF gradient play distinct roles (proliferation and migration, respectively) in new blood vessel formation \cite{Gerhardt2008}.
Cells interact with their microenvironment, which is characterized by the local concentration of astrocyte-derived VEGFA, and evolves according to the following equation:
\begin{equation}\label{eq:VEGF}
\frac{\partial [VEGF]}{\partial t} = D\nabla^{2}[VEGF] + s - \delta [VEGF],
\end{equation}
where $[VEGF]$ is the VEGFA concentration which diffuses in the domain (with no flux boundary conditions), $D$ is the diffusion coefficient, $s$ represent the secretion rate by astrocyte, and $\delta$ the decay rate. Chemotaxis can be incorporated by including an additional reduction in the total energy (equation (\ref{eq:Hamiltonian})) for extensions and retractions towards higher concentrations of VEGFA (as described in \cite{Savill1997})
\begin{equation}\label{eq:chemotaxis}
H'_{chem} = \lambda_{chem}\left(VEGF(\vec{x}) - VEGF(\vec{x}') \right),
\end{equation}
where, $\vec{x}'$ is the neighbor into which site $\vec{x}$ copies its id ($\sigma$), and $\lambda_{chem}$ is the strength of the chemotactic sensitivity to VEGF gradients.

In \cite{Qutub2009,Bentley2008,Bentley2009}, a fixed distribution of VEGFA was specified whereas in our model VEGFA can be dynamic (its concentration can change with position and time). As we will see in the next section, its concentration regulates levels of Delta \cite{Liu2003}, whereas local gradients in its spatial distribution guide tip cell migration and, thereby, sprout polarization.

\FloatBarrier
\subsection*{Subcellular level: modeling lateral-inhibition}
At the subcellular level, tip cell activation is regulated via the Notch-Delta signaling pathway, which is stimulated by VEGFA. The contact lateral-inhibition effect for the exchange of the EC (stalk-tip) phenotype is implemented using a modification of a well established mathematical model that has been proposed by Collier \emph{et al.} \cite{Collier1996}. The model is defined by a set of coupled ordinary differential equations (ODEs), which describe the dynamic processes of
Delta and Notch activation and inhibition between cells that are in contact with each other.
Motivated by the experimental work of Lobov \emph{et al.} \cite{Lobov2007}, where the authors have shown that Delta is induced by VEGFA in the retinal vasculature, we extend Collier's model by incorporating the contribution of VEGFA (as defined in equation (\ref{eq:VEGF})) in the following non-dimensional ODE system,
\begin{eqnarray}
\begin{array}{lll}
\vspace{0.3cm}
Delta:\:\:\:\:\:\:\:\:\:\:\:\:\:\:\:\:\:\:\:\:\:\:\:\:\:\:\dfrac{d D_j}{dt}= v\left( \alpha \dfrac{[VEGF_j]}{VEGF_h + [VEGF_j]}\dfrac{1}{1 + bN^{2}_j} - D_j \right) \;, \\
\vspace{0.3cm}
Notch:\:\:\:\:\:\:\:\:\:\:\:\:\:\:\:\:\:\:\:\:\:\:\:\:\:\dfrac{d N_j}{dt}= \dfrac{\bar{D_j}^2}{a + \bar{D_j}^2} - N_j\;,\\
\vspace{0.3cm}
trans-Delta:\:\:\:\:\:\:\:\:\:\:\bar{D}_j= \sum_{i} \dfrac{D_i P_{ij}}{P_j} \;.\\
\end{array}
\label{eq:ODEmodel}
\end{eqnarray}
where,  $D_j$, $N_j$, are the levels of Delta and Notch in cell $j$, respectively.
$[VEGF_j]=\dfrac{\sum_{i}^{\omega} VEGF_{ji}}{a_j}$ is the average VEGFA in a cell $j$; that is, the sum of VEGFA at every pixel $i$ inside cell $j$ over the cell area, $a_j$, where $\omega$ is the total number of pixels in cell $j$.
$VEGF_h$ is the VEGFA level at which the production rate of Delta is half maximal. The trans-Delta ($\bar{D}_j$) is taken to be the sum over the immediate (contacting) neighbors $i$ of cell $j$. $P_j$ is the perimeter of cell $j$, and $P_{ij}$ is the common area of cell $j$ with its neighbor cells $i$, which is defined as
\begin{equation}\label{eq:ContactArea}
P_{ij} = \sum_{(\vec{x},\vec{x'})}\left(1- \delta_{\sigma(\vec{x}),\sigma(\vec{x'})}\right) \left(1- \delta_{\sigma(\vec{x'}),0}\right) \delta_{\sigma(\vec{x}),i}.
\end{equation}
The summation being over all pairs of adjacent sites in the lattice.

Equations (\ref{eq:ODEmodel}) describe (i) the activation of Notch production within each cell as a function of the levels of (trans-) Delta expressed by neighboring cells, and (ii) the inhibition of Delta expression by Notch. 
The novelty in our model is (iii) the activation of Delta production by extracellular VEGF. In the absence of VEGF signaling, there is no up-regulation of Delta and, therefore, no tip cell activation. For a detailed mathematical analysis on perturbations of the homogeneous steady state of equations (\ref{eq:ODEmodel}) and parameter ranges in which the `salt-pepper' pattern is maintained the reader is referred to Supplementary Information. In the next section results from our multiscale model are presented. We implemented equation (\ref{eq:ODEmodel}) using the Systems Biology Workbench \cite{SBW2006} integrated within the CC3D \cite{ViviAndasari2012}.


\FloatBarrier
\subsection*{Parameters}
The default parameter values used for our simulations are summarized in Table \ref{table:Parameters}, unless otherwise stated. Below we provide a discussion on how some of those were estimated.

\FloatBarrier
\subsubsection*{Cell-cell adhesion: contact energies ($J$)}
Interactions between neighboring pixels have an effective energy, $J$ (as it appears in equation (\ref{eq:Hamiltonian})), which characterizes the strength of cell-cell adhesion. Larger $J$ means more energy is associated with the interface between two cells, which is less energetically favorable, corresponding to weaker adhesivity.

In the simulation results that follow, we use the following set of contact energies: $J_{s,t}=3$, $J_{s,s}=3$, $J_{t,t}=10$, where $s$, $t$ stand for stalk and tip cell, respectively. In doing so, we assume strong adhesion between stalk and tip cells, and weaker adhesion between tip cells. Therefore, fusion of two tip cells will emerge dynamically (from the chemotactic direction and/or ECM alignment), and not from preferential adhesion between tip cells.

\FloatBarrier
\subsubsection*{Tip cell activation}
In Qutub and Popel \cite{Qutub2009}, a tip cell can be activated if and only if the VEGF concentration exceeds a threshold value, [VEGF]$>$0.5 ng/ml.

In our model, a tip cell is activated if its Delta level exceeds a non-dimensional threshold value ($D^*=0.27$). We remark that this threshold has not been evaluated from any experimental measurements, but this may be evaluated in future experiments.


\FloatBarrier
\begin{table}[ht]
\caption{Default parameter settings for simulations} 
\centering 
\begin{tabular}{l l l l } 
\hline\hline 
Parameter & Description & Value & Reference \\ [1ex] 
\hline 
Cell features: & & \\
A                        &  cell target area                             & 100 $\mu m^2$ & \cite{Wiltbank1994} \\ [1ex]
T                        &  cell-membrane fluctuations                   & 1 & / \\ [1ex]
$\lambda_{area}$         &  resistance to changes in size/area           & 10   & / \\ [1ex]
$\lambda_{length}$       &  resistance to changes in length              & 15   & / \\ [1ex]
$\lambda_{chem}$(stalk) &  chemotaxis strength for stalk cells           & 50  &  / \\ [1ex]
$\lambda_{chem}$(tip)   &  chemotaxis strength for tip cells             & 200 & derived \\ [1ex]
$\lambda_{ECM}$          &  strength of preferential attachment to ECM   & 60  & derived  \\ [2ex]
Adhesion: & & \\
$J_{s,t}$                &  stalk-tip contact energy                     & 3    & / \\ [1ex]
$J_{s,s}$                &  stalk-stalk contact energy                   & 3    & / \\ [1ex]
$J_{t,t}$                &  tip-tip contact energy                       & 10   & / \\
[2ex]
VEGFA: & & \\
$D$                      &  diffusion constant VEGFA                     & $10^{-11}$ m$^2$/s & \cite{Bauer2007} \\ [1ex]
$s$                      &  secretion rate VEGFA                         & $0.138$ $\mu$M/s & / \\ [1ex] 
$\delta$                 &  decay rate VEGFA                             & $1.6\textrm{x}10^{-4}$ 1/s  & \cite{Bauer2007} \\
[1ex] 
$D^*$                    &  Delta threshold for tip cell activation      & 0.27 & / \\ [1ex]
Subcellular model: & & \\
$\nu$                   & the ratio of the decay rates of Delta and Notch activities  & 1    & \cite{Collier1996} \\ [1ex]
$\alpha$                & maximal Delta production rate                  & 1 & / \\ [1ex]
$b$                     & strength of Notch negative feedback to Delta   & 100  & \cite{Collier1996} \\ [1ex]
$a$                     & sensitivity of Notch to Delta                  & 0.01 & \cite{Collier1996} \\ [1ex]
$VEGF_h$                & VEGF level at which Delta production rate is half-maximal  & 1    & / \\
\hline 
\end{tabular}
\label{table:Parameters} 
\end{table}

\newpage
\FloatBarrier
\section*{Results}

\FloatBarrier
\subsection*{Experimental results}
\begin{figure}[h!]
\begin{center}
\includegraphics[width=1.0\textwidth,angle=0]{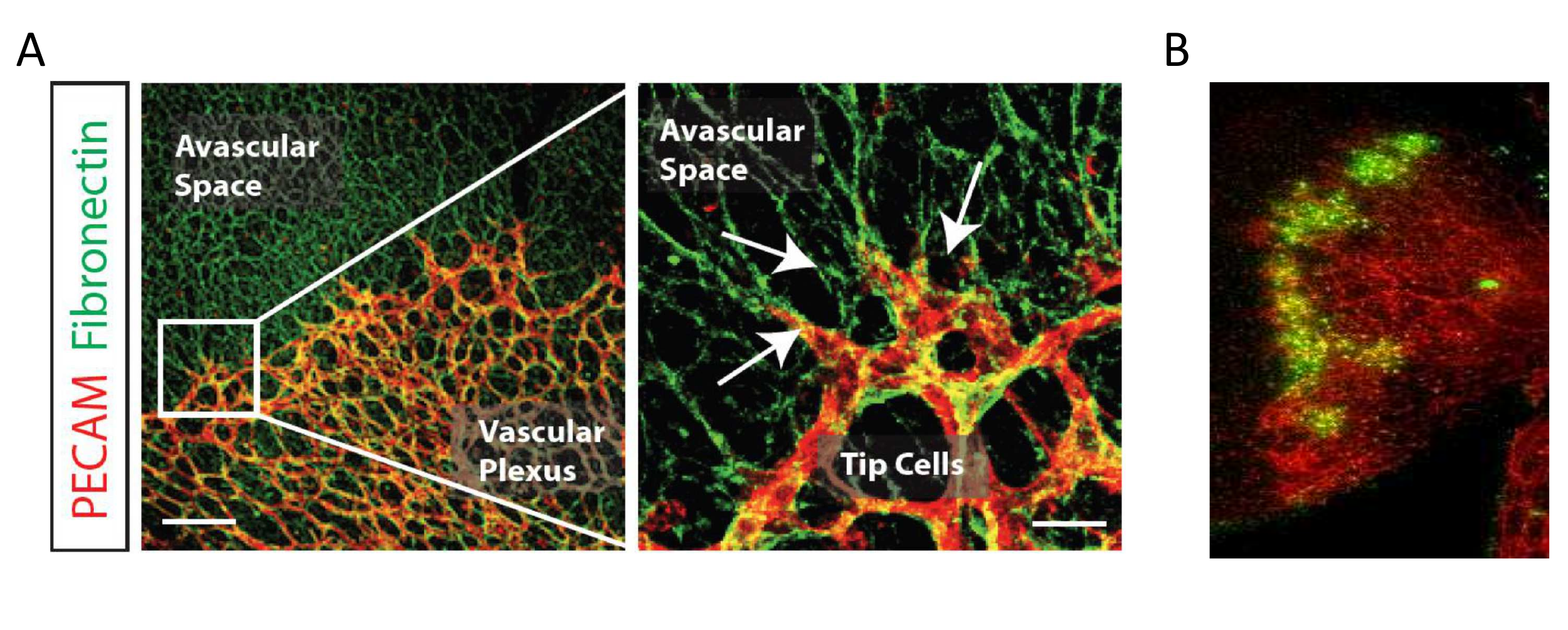}
\label{fig:Luisa_macrophage}
\end{center}
\caption{\textbf{Vascular network expanding over fibronectin-rich matrix.} (A) Blood vessels (PECAM = red)
undergoing active expansion advance over an ECM rich in fibronectin (fibronectin = green).
Scale = 100um. The box has been magnified on the right and demonstrates the intricate association
between tip cells (arrows) and the fibronectin fibrils (green). Scale = 20um. (B) Mouse retina was stained with sFlt1-Fc receptor to detect VEGF (green) and stained with isolectin (red) to label the vascular plexus; scale: 20um. Astrocytes might be the VEGFA source cells \cite{West2005}.}
\label{fig:Luisa_figures}
\end{figure}

In the retina, the vasculature advances centrifugally from the optic nerve to the periphery.  The network of growing capillaries expands on a previously organized template of astrocytes embedded on a fibronectin-rich matrix.  This template of matrix also provides VEGFA that signals to VEGFR2 promoting migration, proliferation and expression of Dll4 by the tip cell.
Figure \ref{fig:Luisa_figures} shows vascular network (red) advancing on a template of fibronectin (green), with the magnified window highlighting single vascular sprouts; the focus of our \emph{in silico} model.

\FloatBarrier

\FloatBarrier
\subsubsection*{Cell cycle kinetics}
\label{sec:cycletime}
\begin{figure}[h!]
\begin{center}
  \includegraphics[width=3in]{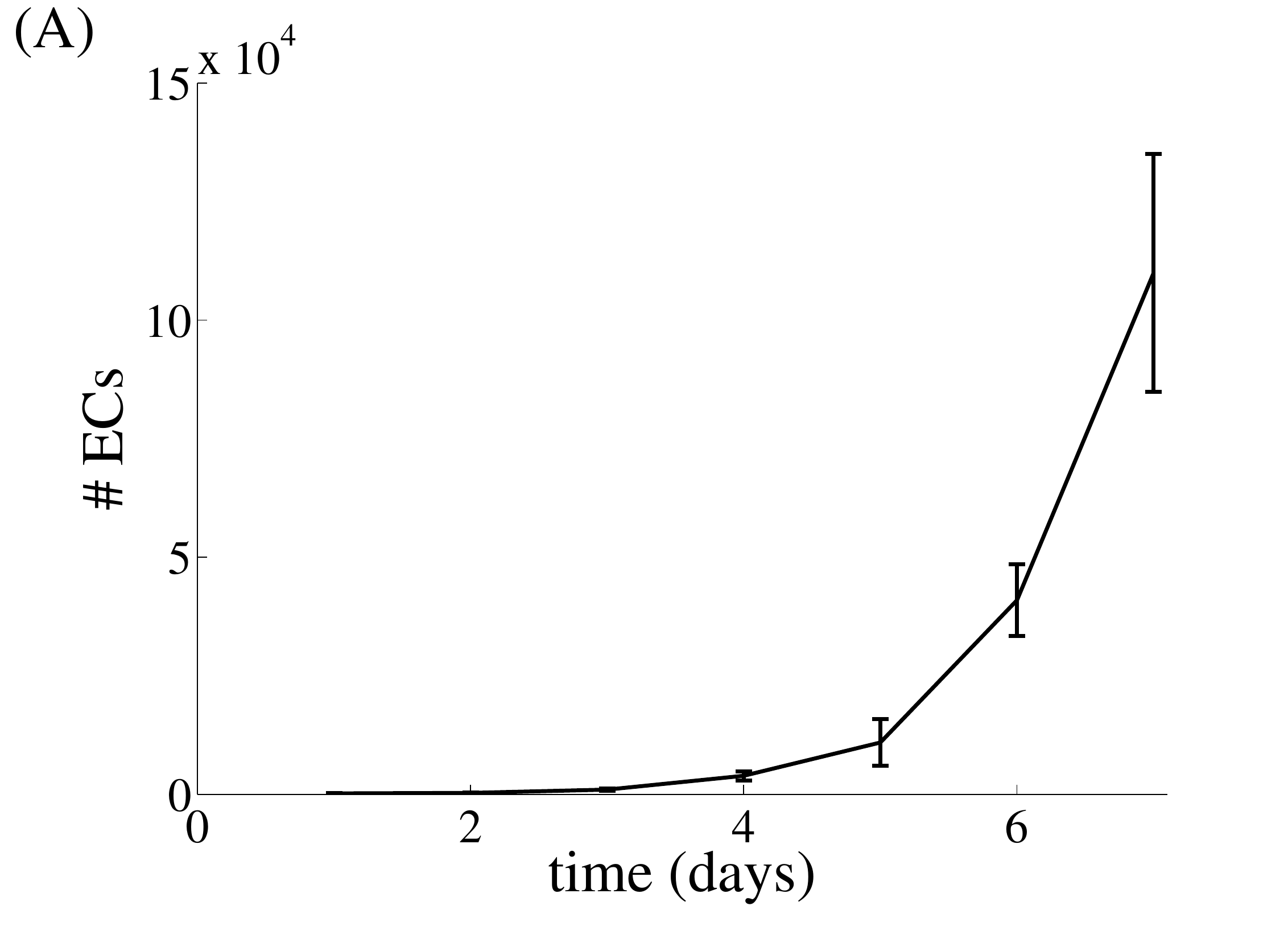} 
\end{center}
\caption{\textbf{Endothelial cells in retina.} Number of endothelial cells from retinas at different developmental (post-natal time) stages. Error bars show the mean $\pm$ S.E.M.}\label{fig:ECs_retina}
\end{figure}

%
%
The proliferation kinetics of the endothelium is an important parameter when modeling the progression of angiogenesis.  The kinetics of cell cycle progression ($t_{cycle}$) was estimated based on data that provided the number of EC from retinas at different developmental (post-natal time) stages, as provided in Figure \ref{fig:ECs_retina}.

EC sprouting from the ophthalmic vein begins around the day of birth, and a dense plexus arrives at the retinal periphery by post-natal day 7.
For this reason we are interested in sprout behavior from day 6 to day 7 (at the end of the angiogenic process) and, therefore, we calculate $t_{cycle}$ as follows,
\begin{eqnarray}\label{eq:cellclock}
  N_7       &=&  N_6 2^{24/t_{cycle}}      \nonumber \\
  t_{cycle} &=&  24 \log_2\left(\frac{N_7}{N_6}\right) \simeq 17 \:\: \textrm{hrs},
\end{eqnarray}
where, $N_j$ is the number of ECs in retina at day $j$.

\FloatBarrier
%
%
%

\FloatBarrier
\subsection*{Computational results}
In this section we consider first the influence of cell shape on Notch-Delta (`salt-pepper') patterning, and then different scenarios regarding VEGF and ECM profiles.

\FloatBarrier
\subsubsection*{Dynamic cell shape}
\label{sec:influenceDNpatterning}
\begin{figure}[h!]
\begin{center}
\includegraphics[width=0.33\textwidth,angle=0]{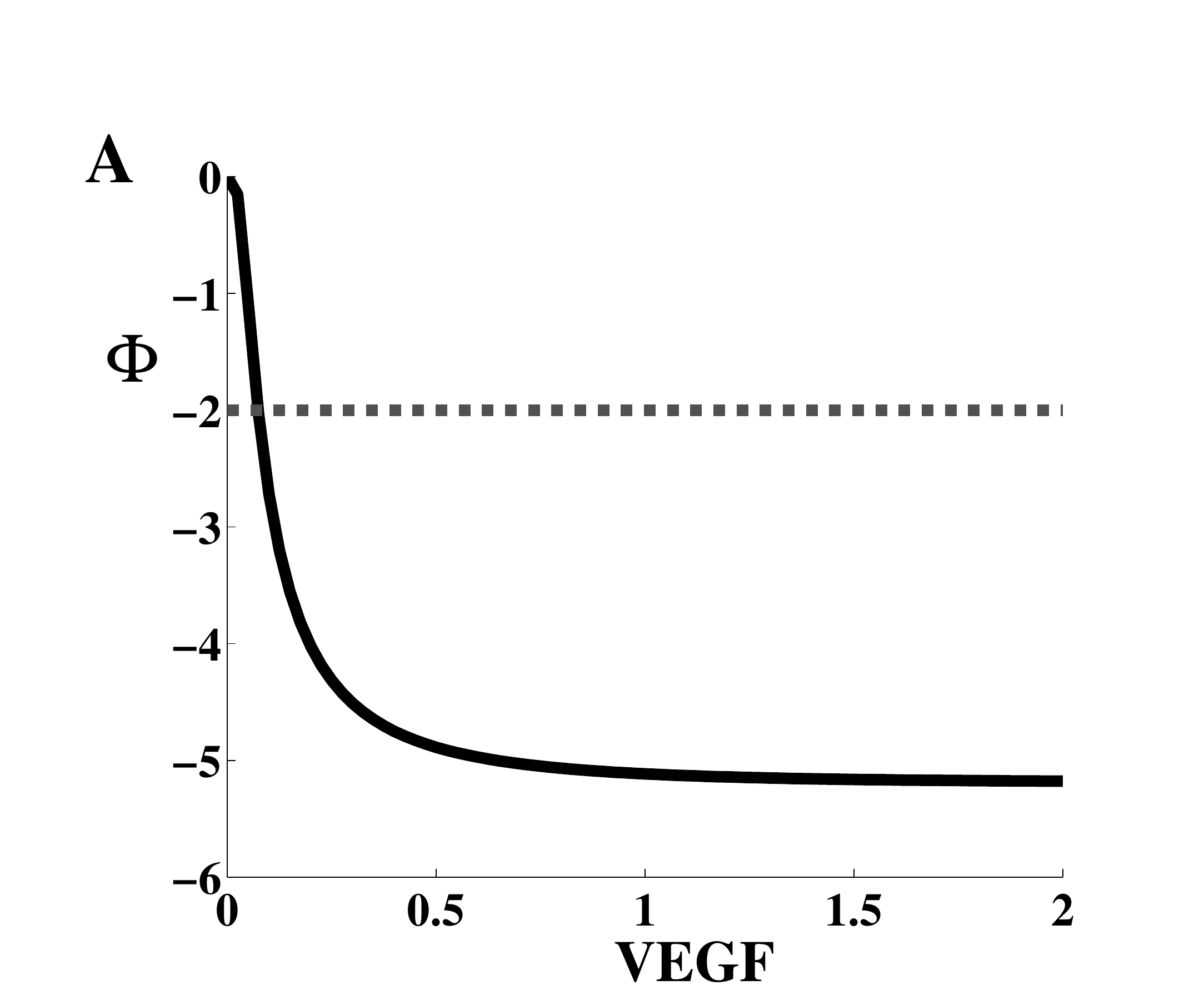}
\label{fig:vary_Js_a}
\includegraphics[width=0.3\textwidth,angle=0]{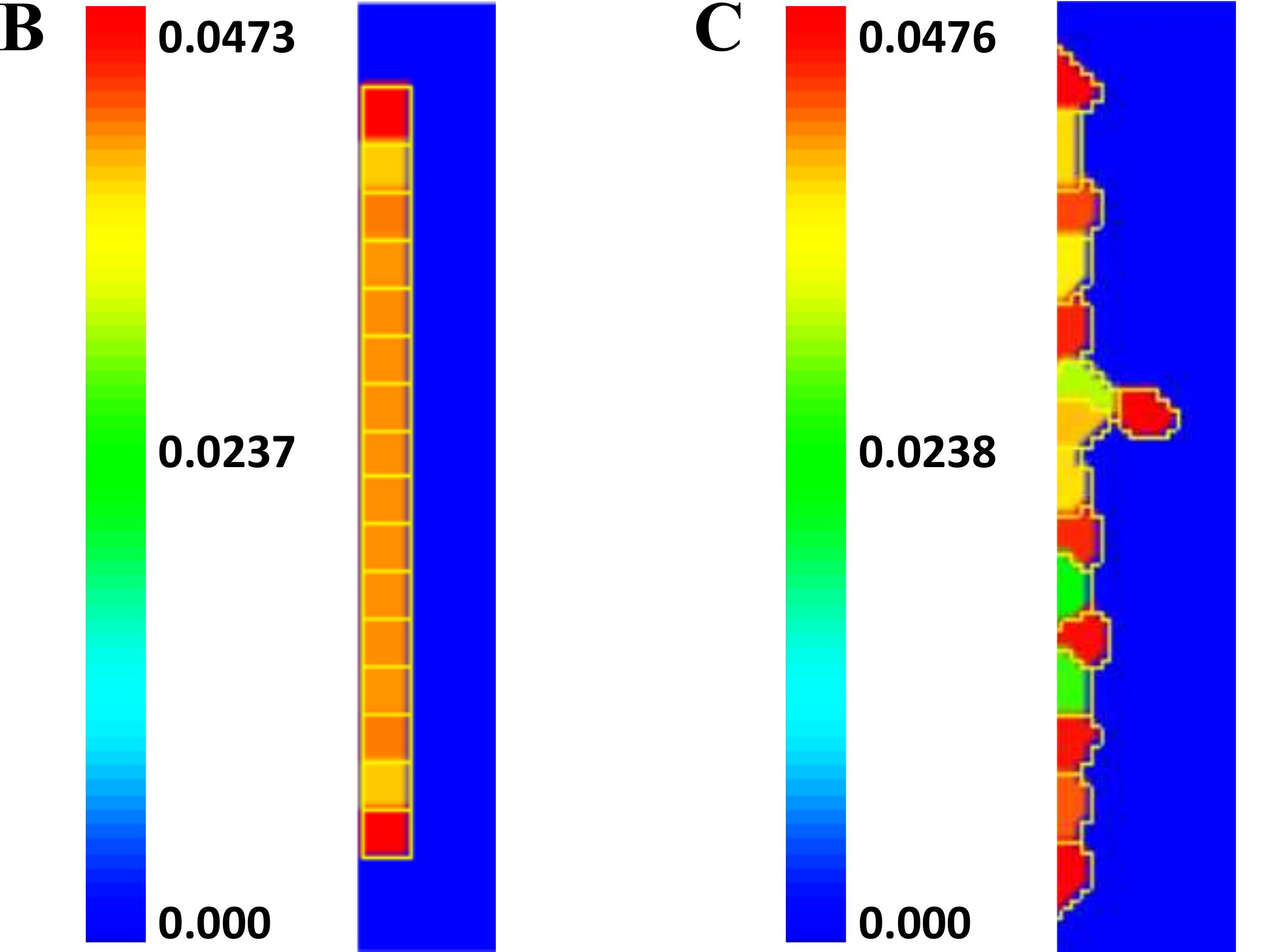}
\label{fig:vary_Js_b}
\end{center}
\caption{\textbf{Influence of cell shape on Notch-Delta patterning.} A) $\Phi$ as in equation (\ref{eq:FG_Collier}) over uniform VEGF. Patterning occurs when $\Phi<-2$ which implies a threshold on VEGF (VEGF$>0.07$). (B-C) Delta profile with and without fixed cell geometries (with/without cell movement), respectively, in 1000 MCS. There is no patterning in fixed cell geometry for VEGF$<0.07$ as expected, however, patterning occurs when all cell behaviors (e.g. membrane fluctuations, elongation, stalk-tip transitions) are allowed except chemotaxis. Parameters for the model (\ref{eq:ODEmodel_new}): VEGF$=0.05$, $\alpha=1$.}
\label{fig:PhiVEGF}
\end{figure}

Figure \ref{fig:PhiVEGF}(A) shows that `salt-pepper' patterning in strings occurs when $\Phi = \alpha h(V)FG < -2$ (see equation (\ref{eq:FG_Collier}) for $\Phi$ and equation (\ref{eq:ODEmodel_FG}) for the definitions of $F$ and $G$) in the Appendix. Briefly, the last inequality implies that $h(V)$ should be sufficiently large (with VEGF$>0.07$), and lateral inhibition $(FG)$ should be sufficiently strong for patterning.
Figure \ref{fig:PhiVEGF}(B) shows that in fixed cell geometries (without cell movement, fluctuations, proliferation or elongation) there is no patterning for VEGF below this threshold.
However, if cells are allowed to move, this condition does not necessarily hold, and we can generate patterning for VEGF$<0.07$ (see Figure \ref{fig:PhiVEGF}(C)). This happens because if cells are not fixed, $\bar{D}_j$ as in equation (\ref{eq:ODEmodel}) changes since $P_{ij}$ (the contact area between cells $j$ and its neighbors $i$) changes. In Figure \ref{fig:PhiVEGF}(b), the cells at the ends of the string have larger Delta levels than the others, since they have only one neighbor and hence a smaller value of $\bar D_j$ (trans Delta), which results in increased Delta.

\FloatBarrier

\FloatBarrier
\subsubsection*{Scenarios regarding VEGF and ECM profile}
Here we investigate the effect of different VEGF and ECM profiles (summarized in Table \ref{table:scenarios}) on sprout morphology and evolution. Figure \ref{fig:incrementaloverview} shows representative snapshots of sprout evolution in each scenario, and below we provide more detailed results.

\FloatBarrier
\begin{table}[ht]
\caption{Different scenarios regarding VEGF and ECM profiles presented in Figure \ref{fig:incrementaloverview}} 
\centering 
\begin{tabular}{l l} 
\hline\hline 
    & \textbf{Scenarios} \\ [1ex] 
\hline 
1.  & homogeneous VEGF $\&$ homogeneous ECM        \\ [1ex]
2.  & homogeneous VEGF $\&$ heterogeneous ECM      \\ [1ex]
3.  & static VEGF gradients $\&$ homogeneous ECM   \\ [1ex]
4.  & static VEGF gradients $\&$ heterogeneous ECM \\ [1ex]
5.  & heterogeneous VEGF $\&$ homogeneous ECM      \\ [1ex]
6.  & heterogeneous VEGF $\&$ heterogeneous ECM    \\ [1ex]
\hline 
\end{tabular}
\label{table:scenarios} 
\end{table}

In the following, for scenarios 1 and 2 which involve spatially uniform VEGF, we choose sufficiently large VEGF ($=1$) so that patterning can occur. In addition to Notch-Delta patterning, scenarios 1 to 6 allow for cell movement and all the other cell behaviors summarized in Table \ref{table:modelmechanisms}.

\subsubsection*{No VEGF gradient (scenarios 1 $\&$ 2)}
In scenarios 1 and 2, there is no VEGF gradient, which implies no sprout polarization.
Figure \ref{fig:incrementaloverview} shows that cell proliferation and elongation are undirected and, therefore,
stalk and tip cells evenly fill the space. This cell behavior results in a reduced EC migration (as discussed later). That was also found and is consistent with experiments reported in \cite{Gerhardt2003,Mitchell2006}, where a spatial gradient in VEGF was removed in the retina, by increasing expression levels of VEGFA in transgenic mouse models. In scenario 2, the addition of a non-uniform ECM has a weak effect. A parameter which could potentially have an effect on that is the $\lambda_{ECM}$ (the preferential attachment of cells to the ECM). That is, if $\lambda_{ECM}$ is large, the cells are more attracted to the ECM fibers. However, we will see later that this could adversely affect (decrease) sprout extension.

\subsubsection*{Static VEGF gradient (scenarios 3 $\&$ 4)}
In these scenarios we incorporate static VEGF gradients, which eventually lead to swollen sprout formation either with or without ECM.

Results in scenarios 1 to 4 approximately up to day 1 look quite similar. That is, sprouts are dominated by single elongated tip cells.
However, there are distinct differences on days 2 and 3. Particularly, in scenarios 3 and 4, cell proliferation is focused on single sprouts as a result of the steep VEGF gradients.

\subsubsection*{Dynamic VEGF from single source (scenarios 5 $\&$ 6)}
Here, a fixed astrocyte (VEGF source) is responsible for the VEGF gradients.
The resulting morphology of the capillary sprout is determined by two main mechanisms: the astrocyte-derived VEGFA, which activates the Delta activity in each cell (according to equations (\ref{eq:ODEmodel})), and the Notch-Delta signaling pathway which yields the `salt-pepper' pattern
with tip cells migrating up VEGFA gradients.

\begin{figure}[h!]
\begin{center}
  \includegraphics[width=6.5in]{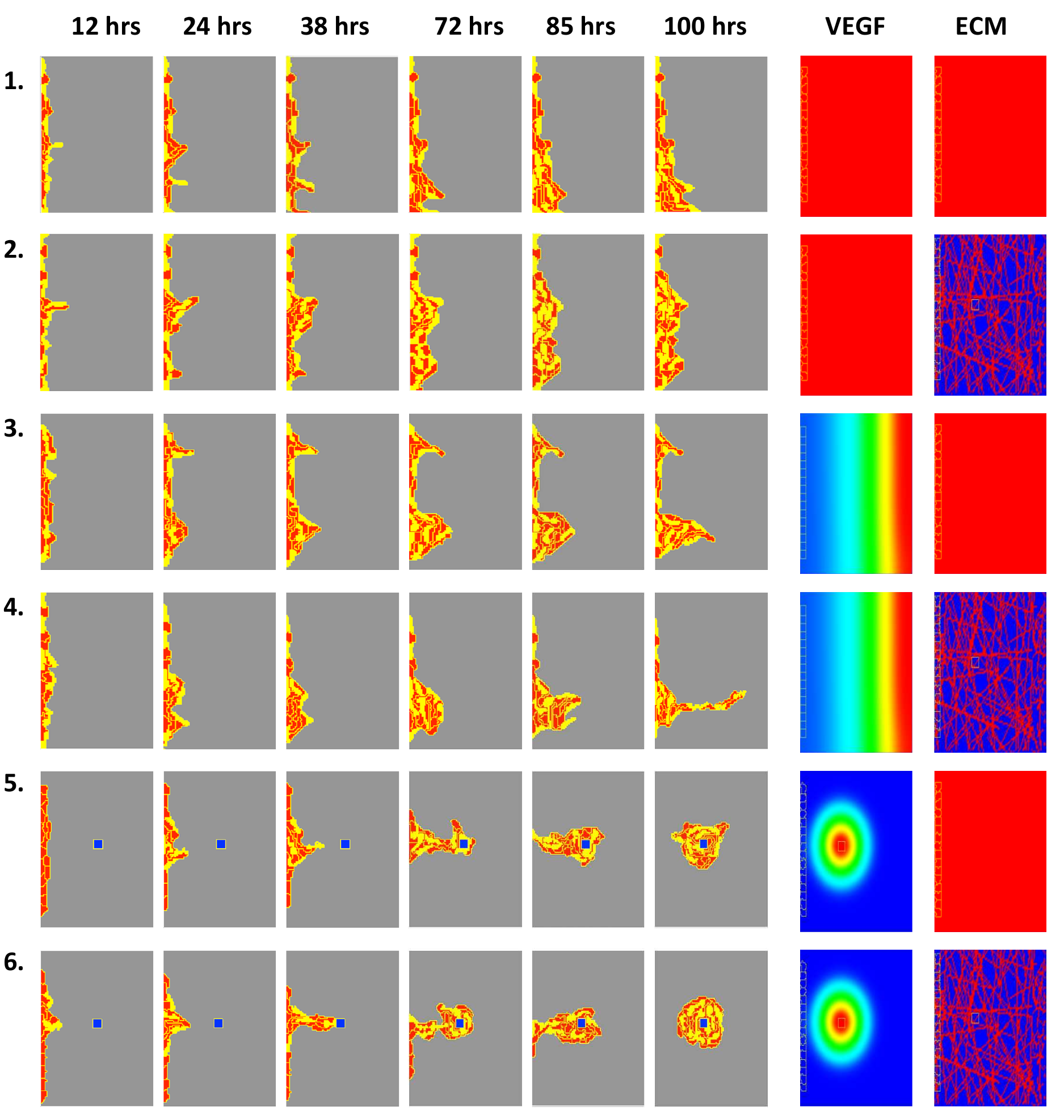} 
\end{center}
\caption{\textbf{Sprout evolution in different VEGF and ECM profiles.}
Representative simulation snapshots (from 10 simulations) of sprout evolution for the six scenarios outlined in Table \ref{table:scenarios}. Scenario 6 appears at 18 hrs to be the closest approximation... Key: stalk cells (red), tip cells (yellow), fixed astrocyte (blue).
}\label{fig:incrementaloverview}
\end{figure}

Figure \ref{fig:incrementaloverview} (scenarios 5 and 6) demonstrates the model's ability to reproduce realistic capillary sprout morphologies (up to $\sim$18 hrs).
Scenario 5 (with homogeneous ECM) can give a polarized sprout, but scenario 6 with heterogeneous ECM gives narrower sprouts in 12-18 hrs.

Therefore, we suggest that scenario 6 provides a close approximation to a growing vascular sprout. However, since the astrocyte is fixed it does not allow for a longer sprout formation, because at late time points (days 2-3) we observe a mass of cells surrounding the astrocyte.

\newpage
\FloatBarrier

\newpage
\break
\FloatBarrier
\subsubsection*{Extension speed matches experiment with dynamic VEGF and presence of ECM}
As we described before, scenario 6 provided the best approximation to sprout morphology as in Figure \ref{fig:schematic_sprout_morphology},
and henceforth, all results which proceed are derived under this scenario, unless otherwise stated.

Sprout length is determined by measuring the distance from the parent blood vessel (located at the one side of the domain) to the leading tip cell's centre of mass. Average sprout extension speed is then calculated as the final sprout length over time. Simulation results (up to 18 hrs; the time needed from the sprout to reach the astrocyte in scenario 6) showed that scenarios 5 and 6 are the ones with the highest speed and in agreement with experimental data ($\sim$3.5 $\mu m/hr$) since cell proliferation is focused on a single sprout, whereas the speed in scenarios 1-4 is much slower ($\sim$1 $\mu m/hr$).

\begin{figure}[h!]
\begin{center}
\includegraphics[width=0.235\textwidth,angle=0]{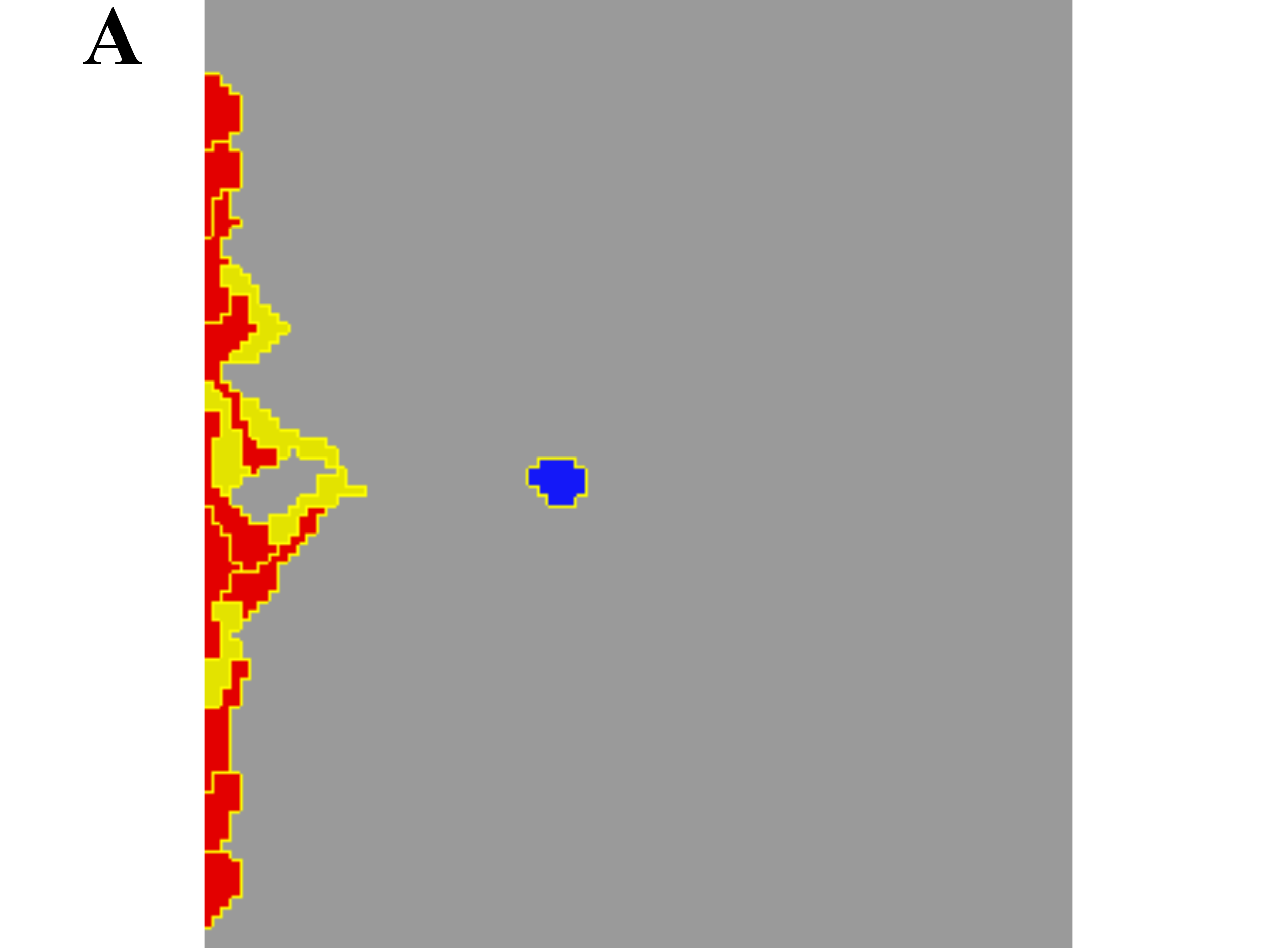}
\label{fig:spr_morph_3}
\includegraphics[width=0.235\textwidth,angle=0]{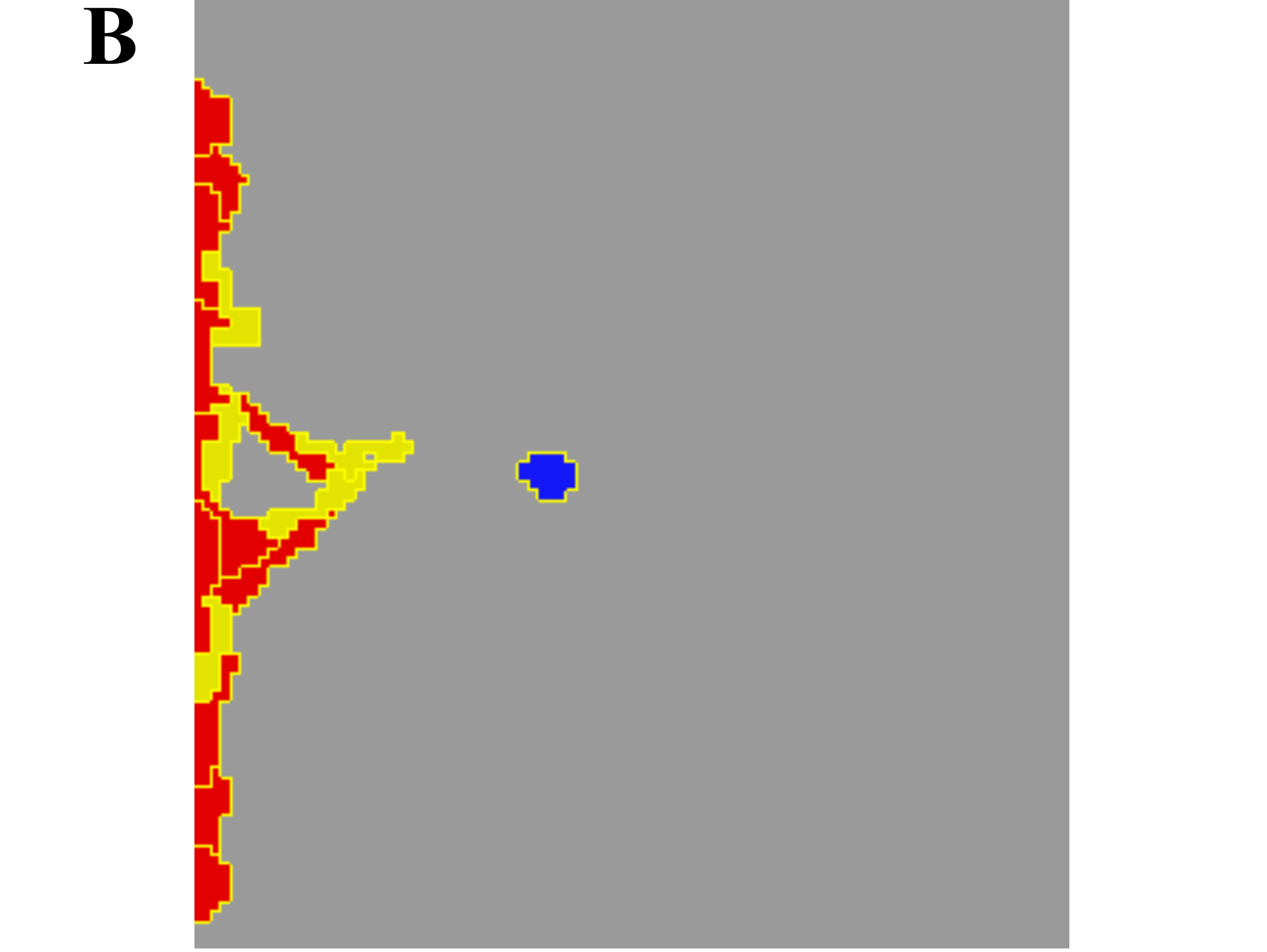}
\label{fig:spr_morph_4}
\includegraphics[width=0.235\textwidth,angle=0]{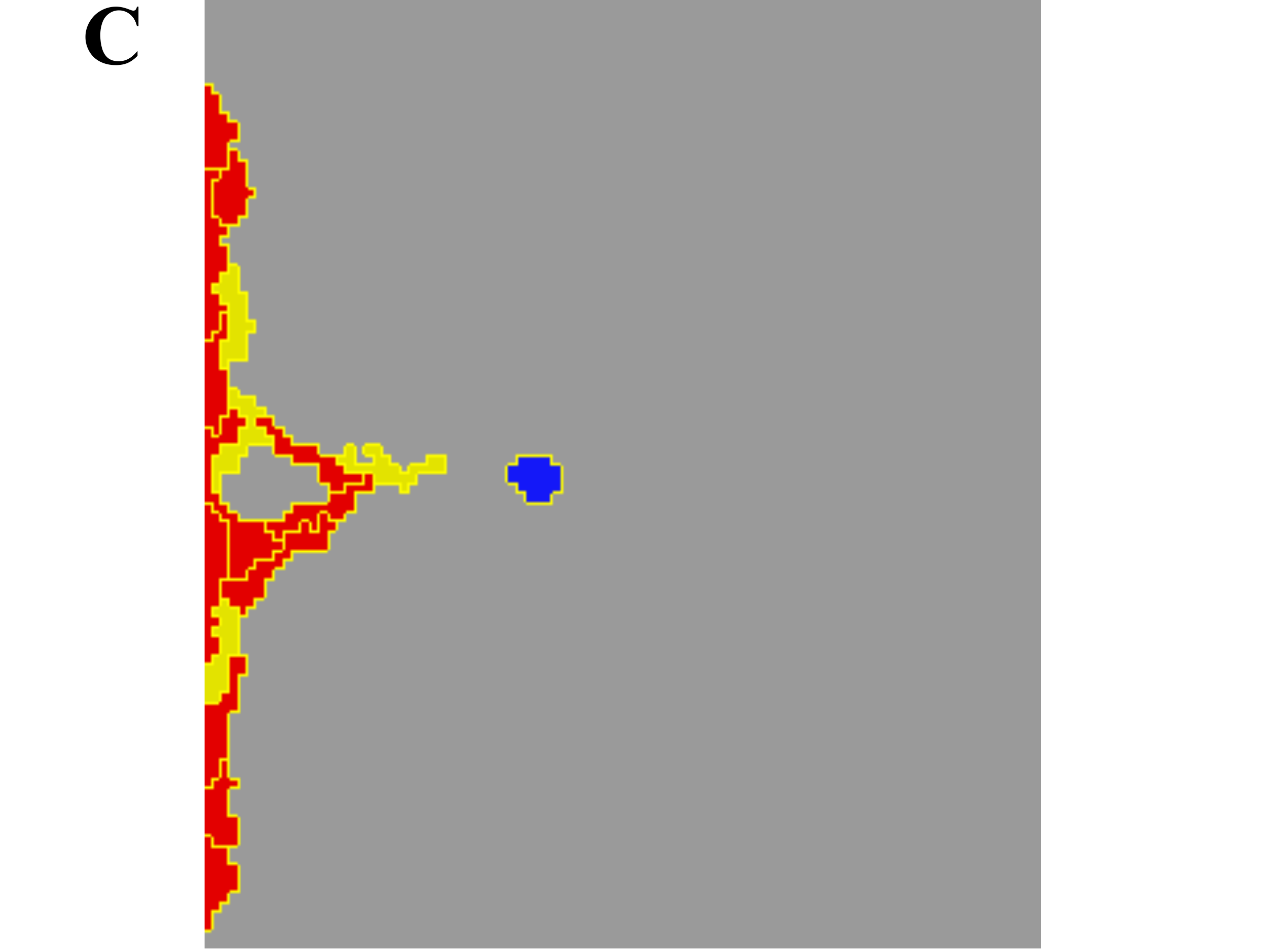}
\label{fig:spr_morph_5}
\includegraphics[width=0.235\textwidth,angle=0]{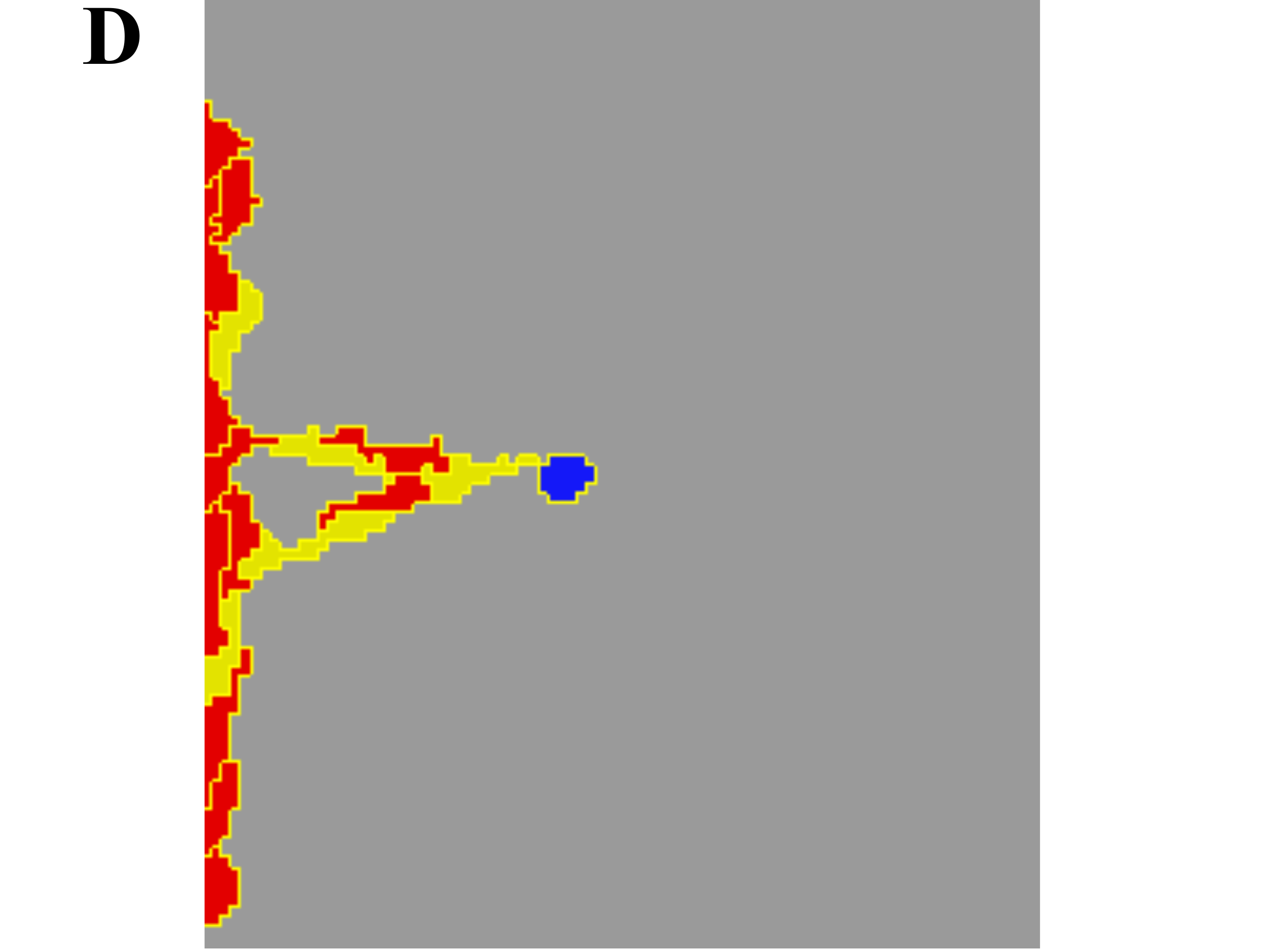}
\label{fig:spr_morph_6}
\end{center}
\caption{\textbf{Sprout anastomosis (fusion).} Representative simulation snapshots from scenario 6 (different realization from Figure \ref{fig:incrementaloverview}) of sprout evolution showing the model's ability to reproduce anastomosis. (A)-(B) Tip cell fusion (two adjacent tip cells) in 8 and 12 hrs, (C) one of the two tip cells becomes a stalk cell (lateral-inhibition effect from Notch-Delta signaling) in 13 hrs, and (D) the leading tip cell moves up the astrocyte-derived VEGFA gradients in 18 hrs. Key: stalk cells (red), tip cells (yellow), astrocyte (blue).}
\label{fig:spr_morph}
\end{figure}

Figure \ref{fig:spr_morph} shows that when the head tip cell of a growing sprout contacts other sprout (other head tip cell), new cell-cell junctions are established and the sprouts become connected (a behavior called anastomosis). 

\FloatBarrier

We also investigated the effect of cell elongation on the sprout formation in scenario 6, and results (not presented here) showed that if stalk cells adjacent to tip cells are not able to elongate, then the sprout splits. Similarly, if both stalk and tip cells are not able to elongate, the same is true. Together this suggests that cell elongation is necessary for sprouting. However, if cell proliferation was increased that could sometimes prevent cell detachment, but the sprout width would increase which is not desirable.

\FloatBarrier
\subsubsection*{Perturbation experiments: Notch-Delta knockouts}
The VEGFA-Dll4-Notch1 signaling appears to be critical to vascular development. In this section we summarize published experimental results on VEGFA-Dll4-Notch1 signaling, and aim to address its importance in sprout growth in \emph{in silico} knockout experiments.

\vspace{0.5cm}

\FloatBarrier
\textbf{\emph{In vitro} knockout experiments of Notch-Delta signaling:}\\
The dynamic interaction between VEGF and Notch-Delta signaling was unravelled independently by several groups while studying the process of angiogenic sprouting in the postnatal mouse retina \cite{Hellstrom2007,Lobov2007,Suchting2007}. They all observed Dll4 expression in ECs at the leading front of the vascular plexus and found that inhibition of Notch-Delta signaling results in increased vascular density due to excessive sprouting.

Similar findings were described in the zebrafish intersegmental vessel (ISV) sprouting model. Inhibition of the Notch pathway induced hyperbranching of the ISVs and leaded to an increased number of ECs \cite{Leslie2007,Siekmann2007}. By contrast, overexpression of the activated Notch receptor blocks sprouting of the ISVs \cite{Siekmann2007}.

In addition, haploinsufficiency (when in diploid organisms one of the two copies of a gene is mutated) of the Dll4 gene was embryonically lethal in many mouse strains as a result of extensive vascular defects \cite{Gale2004,Krebs2004}.

\vspace{0.5cm}

\FloatBarrier
\textbf{Published \emph{in silico} knockout experiments:}\\
Here we briefly discuss results obtained from two \emph{in silico} models which attempted to address the effects of Notch-Delta (Dll4+/-) knockout experiments on vessel development.

Qutub and Popel \cite{Qutub2009} studied the effects of VEGF protein concentrations and Dll4 haploinsufficiency (Dll4+/-) on capillary network formation. Without VEGF or with local VEGF levels less than 0.6 ng/ml, the ECs were not activated. Regarding the effect of haploinsufficiency of Dll4 on blood vessel sprouting, it was found that in Dll4+/- condition, the relative total vessel length and the number of tip cells in 24 hrs increased compared to control conditions. The activity (and haploinsufficiency) of Dll4 in \cite{Qutub2009} is ingrained in the model in a rule-based way, e.g. tip cell proliferation, the maximum number of tip cells, and the rules in branching are different in Dll4+/- and Dll4+/+ conditions.

Bentley \emph{et al.} \cite{Bentley2008} implemented the Dll4+/- \emph{in silico} experiments by varying a parameter $\delta$, which controls the expression level of Dll4 in response to VEGF (equivalent to parameter $\alpha$ in our model (\ref{eq:FG_Collier})).
In particular, a Dll4+/- heterozygous knockout genotype was simulated by halving $\delta$, in which case the lateral-inhibition effect (stalk-tip cell patterning) from Notch-Delta signaling was not possible
The authors suggested that a Dll4+/- mutant would perform normal patterning in twice the VEGF level.

\vspace{0.5cm}

\begin{figure}[h!]
\begin{center}
\includegraphics[width=0.45\textwidth,angle=0]{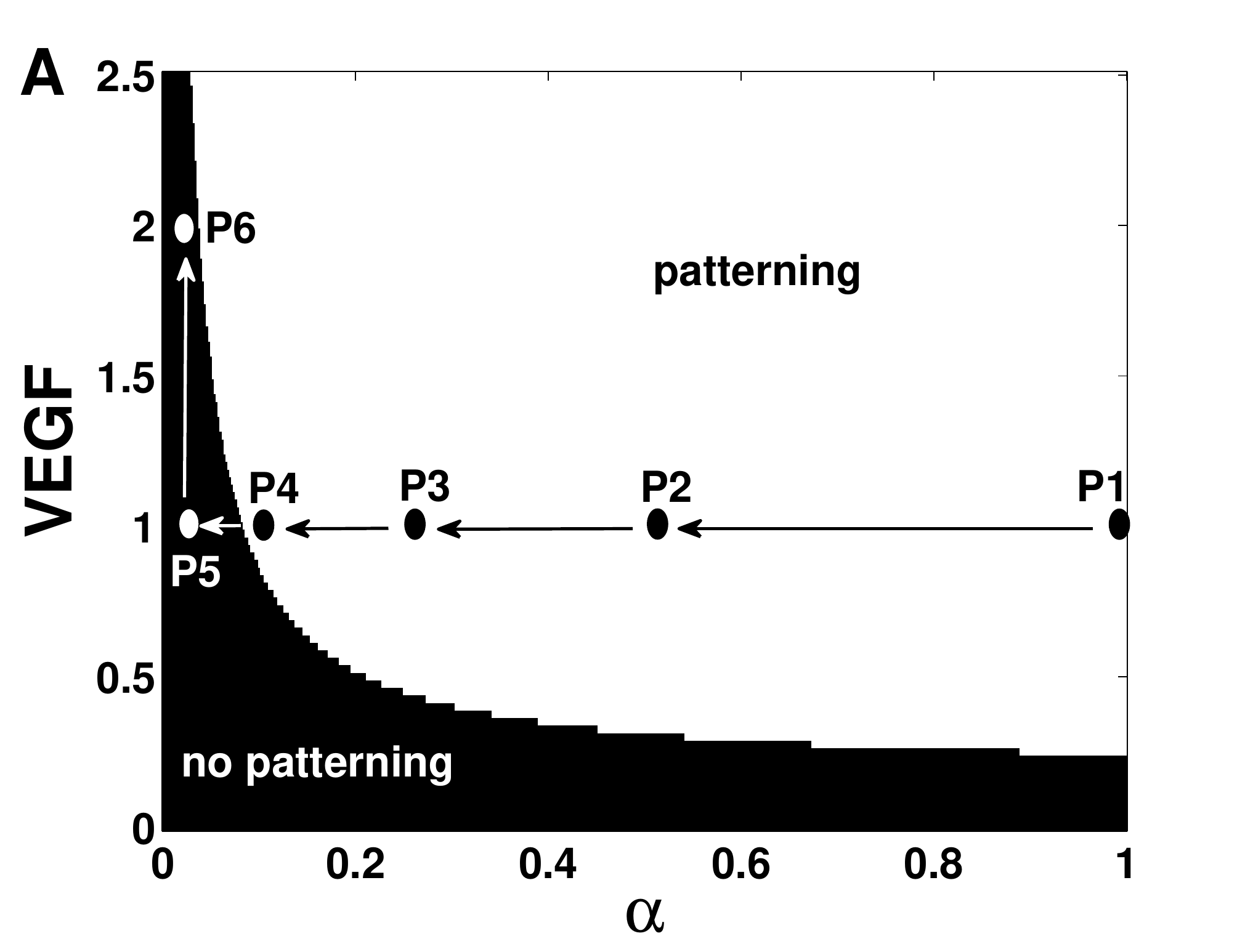} 
\label{fig:spr_morph_4}\\
\includegraphics[width=0.95\textwidth,angle=0]{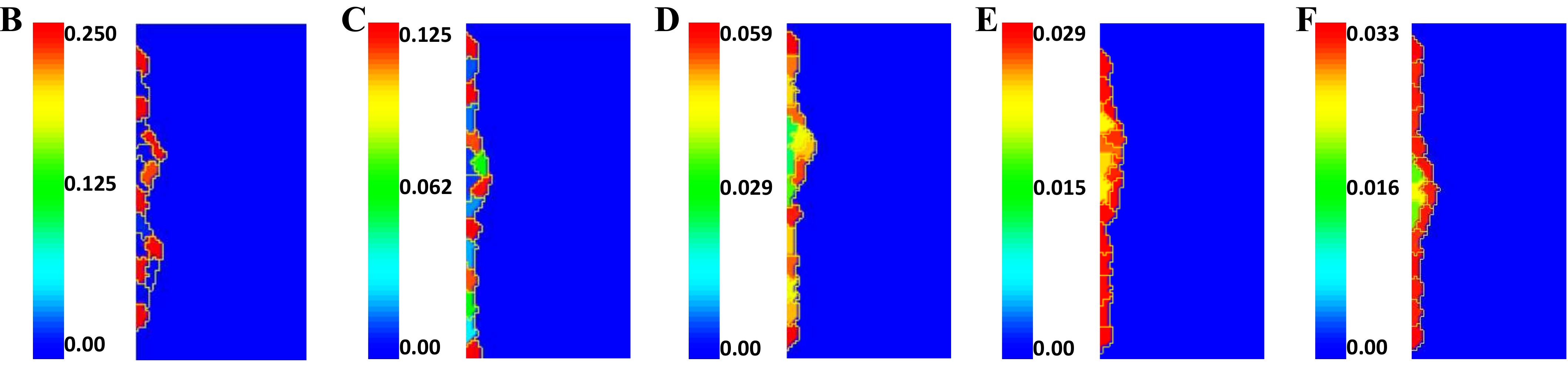}
\label{fig:spr_morph_1}
\end{center}
\caption{\textbf{`Salt-pepper' patterning along the blood vessel in \emph{in silico} Notch-Delta (Dll4+/-) knockout experiments in uniform VEGF.}
(A) VEGF over $\alpha$ (the sensitivity of Delta to VEGF) shows the ranges of `salt-pepper' patterning when $\Phi<-2$ as in equation (\ref{eq:FG_Collier}). The movement from point P1 to P5 (by halving $\alpha$ each time) represents Dll4+/- haploinsufficiency condition. (B)-(F) Delta levels in different combinations of uniform VEGF and $\alpha$ where patterning does or does not occur. In particular, (B) P2; $\alpha=0.5$, (C) P3; $\alpha=0.25$, (D) P4; $\alpha=0.12$, (E) P5; $\alpha=0.06$, with VEGF=1, and (F) P6; $\alpha=0.06$, with VEGF=2.
We observe that in (B) there is a distinct `salt-pepper' pattern (alternating of low and high Delta among neighboring cells). In particular, cells with high Delta are more likely to become tip cells, and cells with low Delta to become stalk cells. However, while $\alpha$ decreases (from (B) to (F)) it appears that adjacent cells do not inhibit each other fully, and thereby, it may not lead to a `salt-pepper' pattern. In the case where $\alpha$ is very low, cells share approximately the same Delta levels, and thereby, they could become either `all stalk' or `all tip' cells (pathological angiogenesis) depending on the Delta threshold for tip cell activation (given in Table \ref{table:Parameters}).}
\label{fig:DN_blockade_2}
\end{figure}

\begin{figure}[h!]
\begin{center}
\includegraphics[width=1.0\textwidth,angle=0]{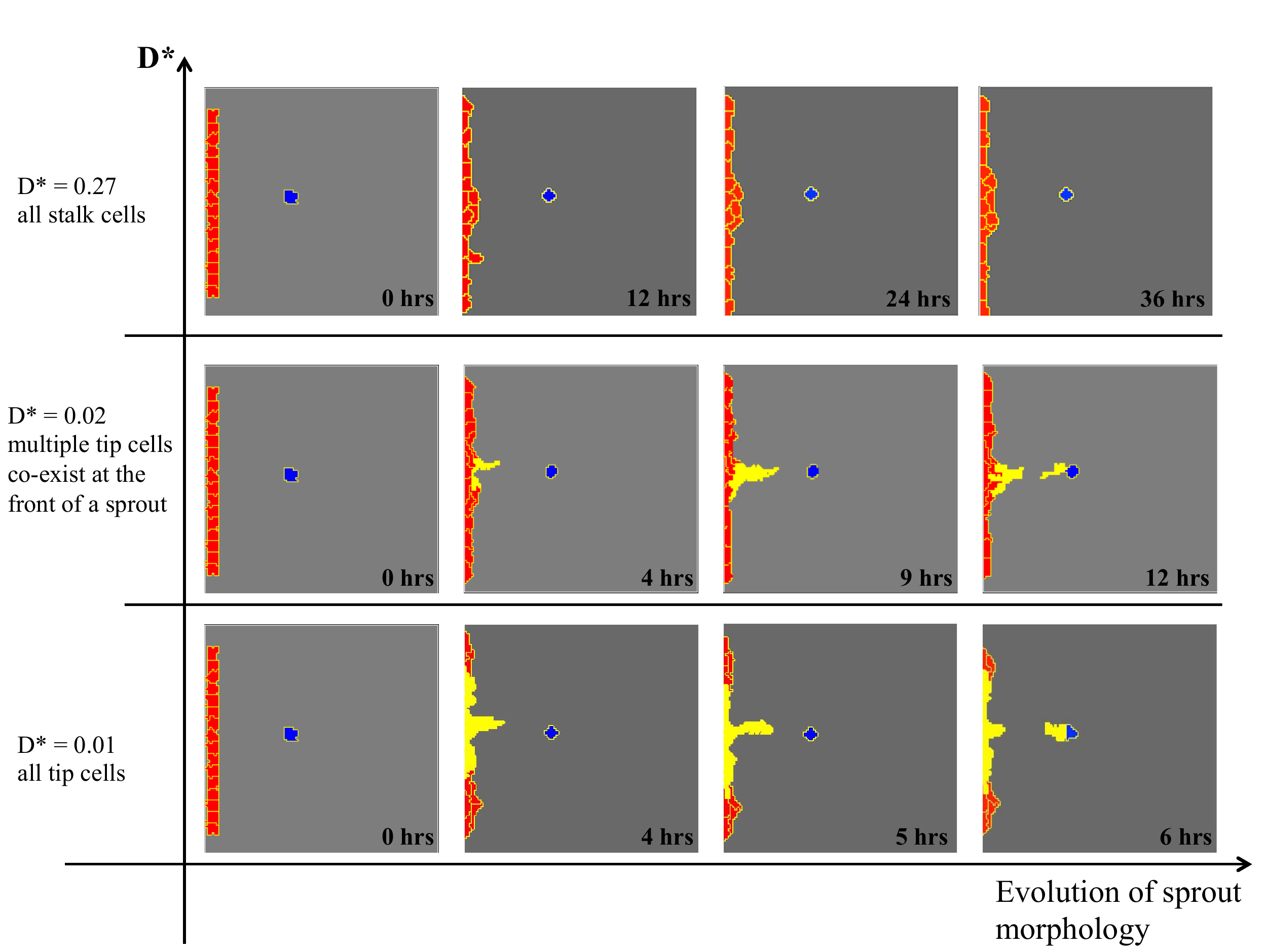}\\
\includegraphics[width=0.5\textwidth,angle=0]{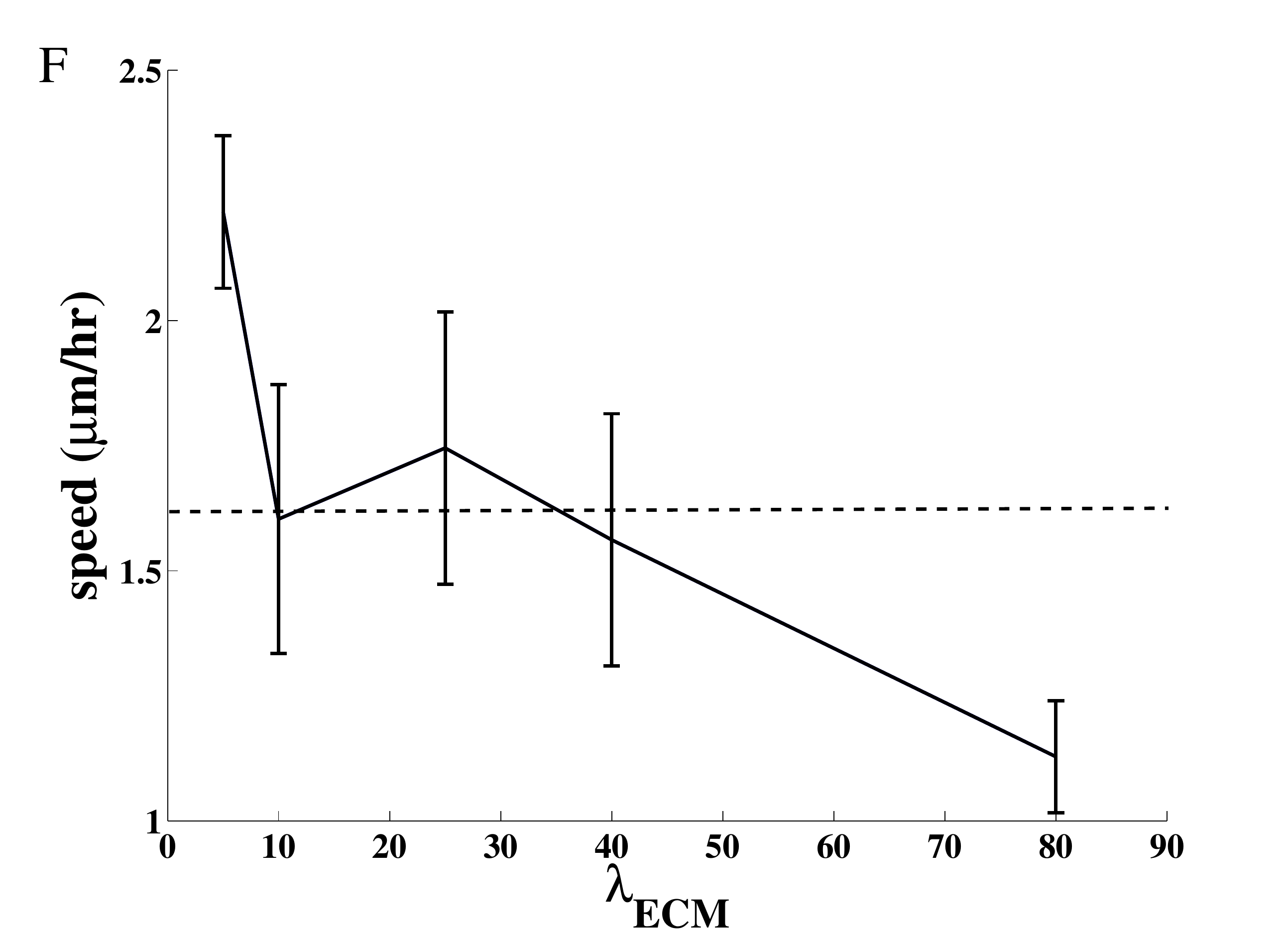}
\end{center}
\caption{\textbf{Sprout formation and extension speed in \emph{in silico} Notch-Delta (Dll4+/-) knockout experiments in non-uniform VEGF by varying $D^*$, the Delta threshold for tip cell activation.}
(A) Evolution of sprout morphology over $D^*$ when $\alpha$, the sensitivity of Delta to VEGF (as in equation (\ref{eq:FG_Collier})), is low (see Figure \ref{fig:DN_blockade_2}(F)). In particular, when $D^*$ is low all cells along the sprout become tip, and their strong chemotaxis towards the astrocyte result in splitting of the parent blood vessel. For intermediate values of $D^*$ the sprout splits due to the accumulation of more than one tip cell at the front of the sprout. Finally, when $D^*$ is high all cells become stalk which result in a thick sprout. Key: stalk cells (red), tip cells (yellow), astrocyte (blue).
(B) Average sprout extension speed from 10 simulations in 7 hrs for $D^*=0.0095$, 13 hrs for $D^*=0.021$, and 42 hrs for $D^*=0.27$. The dashed line corresponds to the average extension speed ($\sim$3.5 $\mu m/hr$) from our experimental data at postnatal day 7. In the case where all cells are stalk, the sprout extension speed is slower than the normal (Dll4+/+) situation. However, the extension speed increases dramatically as $D^*$ decreases. Error bars show the mean of simulations $\pm$ S.E.M.}
\label{fig:DN_blockade_3}
\end{figure}

\FloatBarrier
\textbf{New knockout (Dll4+/-) experiments in our \emph{in silico} model:}\\
\textbf{(A)The impact on `salt-pepper' patterning in uniform VEGF environment}\\
In Figure \ref{fig:DN_blockade_2} we present results from our \emph{in silico} model in a Dll4+/- condition and its impact on `salt-pepper' (alternating stalk-tip cell) patterning in uniform VEGF environment. We follow a similar approach as in \cite{Bentley2008} in order to allow comparison between the two models; that is, vary $\alpha$ (the sensitivity of Delta to VEGF).
Our parameter choice (e.g. VEGF=1 (as in scenarios 1 and 2 in Figure \ref{fig:incrementaloverview}), and $\alpha=1$) in the parameter space ($\alpha$, VEGF) sets our case to be at point P1 (see Figure \ref{fig:DN_blockade_2}(A)). If we move from P1 to P2, or P2 to P3, or P3 to P4 by halving $\alpha$ each time (see Figures \ref{fig:DN_blockade_2}(B)-(D)), then we get patterning for all cases since we are in the white region of Figure \ref{fig:DN_blockade_2}(A), where $\Phi<-2$ as in equation (\ref{eq:FG_Collier}). However, if we move from P4 to P5, we lose patterning (high Delta in all cells). As we mentioned above, Bentley \emph{et al.} in \cite{Bentley2008} suggested that if we double the VEGF level, patterning could be recovered. We show here an example where if we move from P5 to P6 (by doubling VEGF from 1 to 2) patterning is not recovered (see Figures \ref{fig:DN_blockade_2}(E)-(F)).
Therefore, our results suggest that recovery from Dll4 haploinsufficiency by doubling VEGF is not necessarily possible.
\\\\
\textbf{(B)The impact on sprout formation in non-uniform VEGF environment}\\
In Figure \ref{fig:DN_blockade_3} we present results in a Dll4+/- condition and its impact on sprout formation and extension speed in the presence of an astrocyte (non-uniform VEGF environment). We focus on the case where $\alpha$ is low (e.g. $\alpha=0.05$ as in Figure \ref{fig:DN_blockade_2}(F)) where the `salt-pepper' patterning breaks down. In particular, depending on $D^*$, the Delta threshold for tip cell activation (see Table \ref{table:Parameters}), cells tend to become either `all stalk' or `all tip' cells. It is shown that when $D^*$ is high (e.g. $D^*=0.27$) cells become `all stalk'. In this case a thick sprout evolves with a low extension speed (lower from the average extension speed in normal situation; see Figure \ref{fig:DN_blockade_3}(B)) since stalk cells chemotact towards the astrocyte less strongly compared to tip cells. For intermediate values of $D^*$ (e.g. $D^*=0.021$) more than one tip cell start to accumulate at the front of the sprout, which result in the splitting of the new sprout. Finally, when $D^*$ is low (e.g. $D^*=0.0095$) cells become `all tip'. Tip cells strongly chemotact towards astrocyte, which in turn adversely affect the integrity of the parent blood vessel. In conclusion, all these cases characterize pathological angiogenesis, which all emerged from low $\alpha$.

\FloatBarrier

\FloatBarrier


\subsubsection*{Sensitivity analysis}
\label{sec:sensitivityanalysis}
%
We have reported a rather complicated model that considers effects of intracellular Notch-Delta signaling, cell-cell adhesion, and cell-environmental interactions (chemotaxis and haptotaxis). The model has many parameters, listed in Table \ref{table:Parameters}, not all of which have direct experimental measurements. We examine the effects of the parameters whose values we cannot obtain from experiments and study their effects on the simulation results.
In our model, simulations were performed in which parameters were varied independently from their reference values
presented in Table \ref{table:Parameters}.

\vspace{0.5cm}

\FloatBarrier
\textbf{VEGF decay rate:}\\
The spatial distribution of VEGF has several influences on angiogenesis. Firstly, we varied VEGF gradient, via the decay rate of VEGF ($\delta$) as in equation (\ref{eq:VEGF}).

\begin{figure}[h!]
\begin{center}
\includegraphics[width=0.23\textwidth,angle=0]{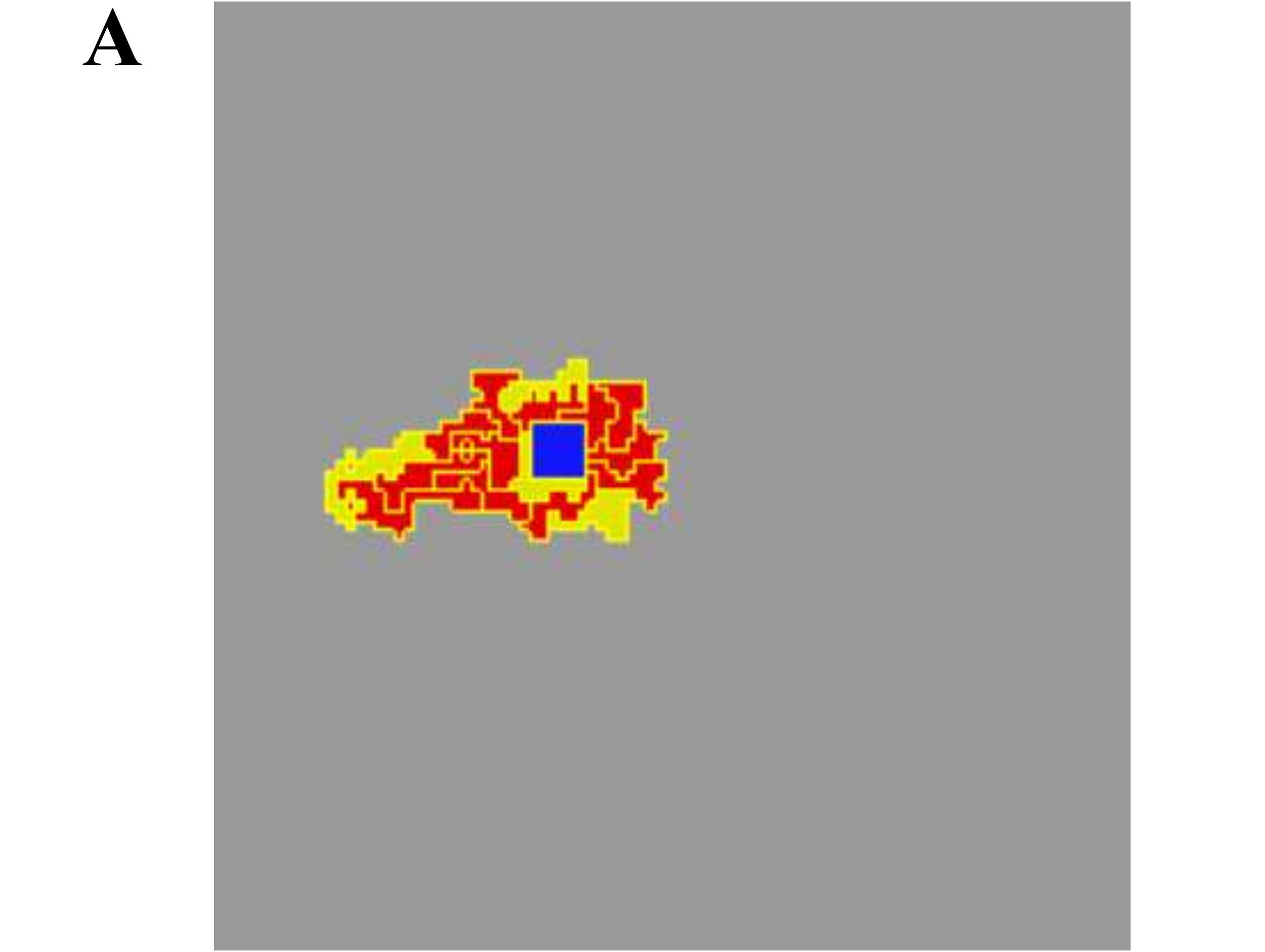} 
\label{fig:spr_morph_1}
\includegraphics[width=0.23\textwidth,angle=0]{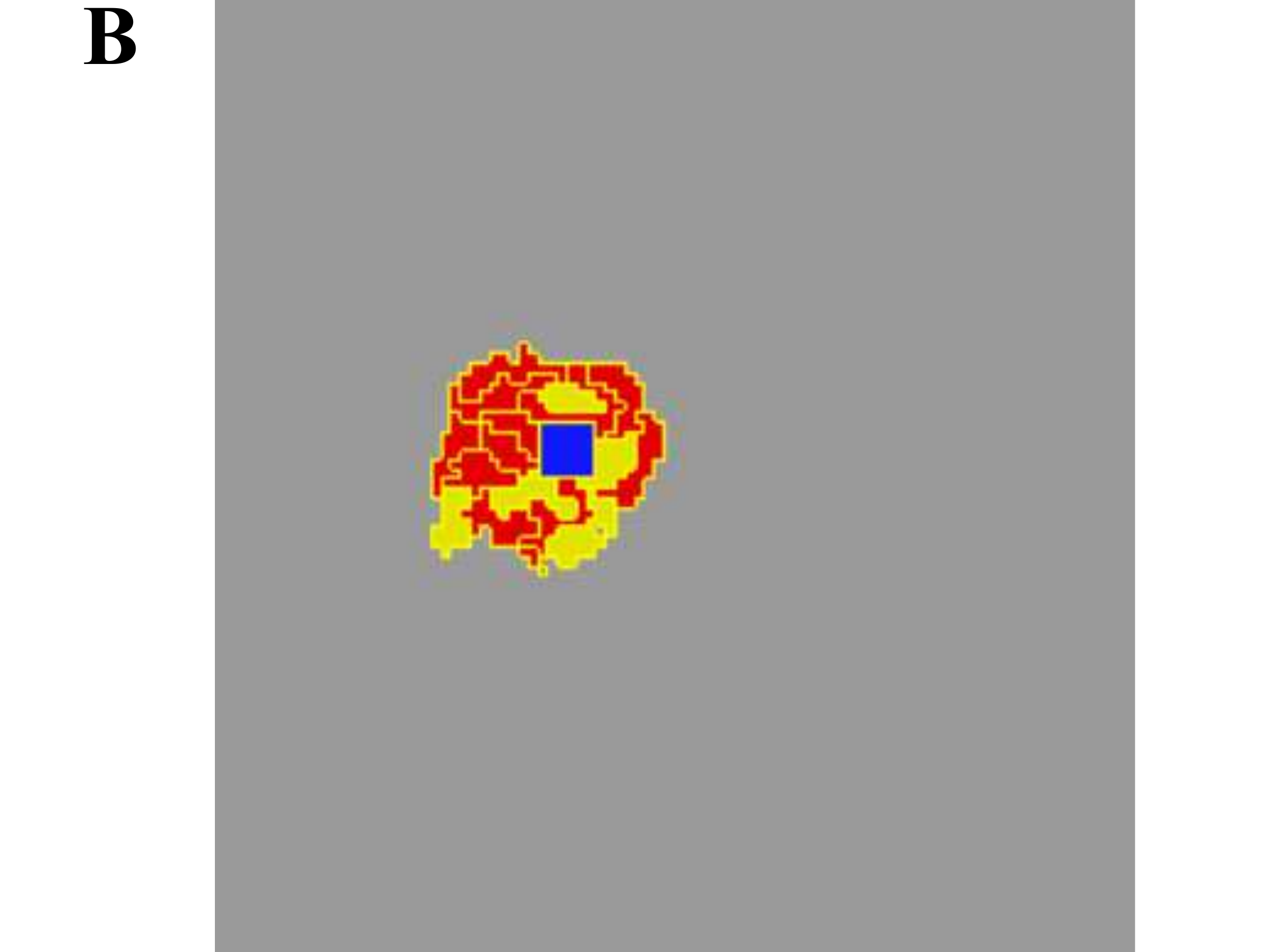} 
\label{fig:spr_morph_2}
\includegraphics[width=0.23\textwidth,angle=0]{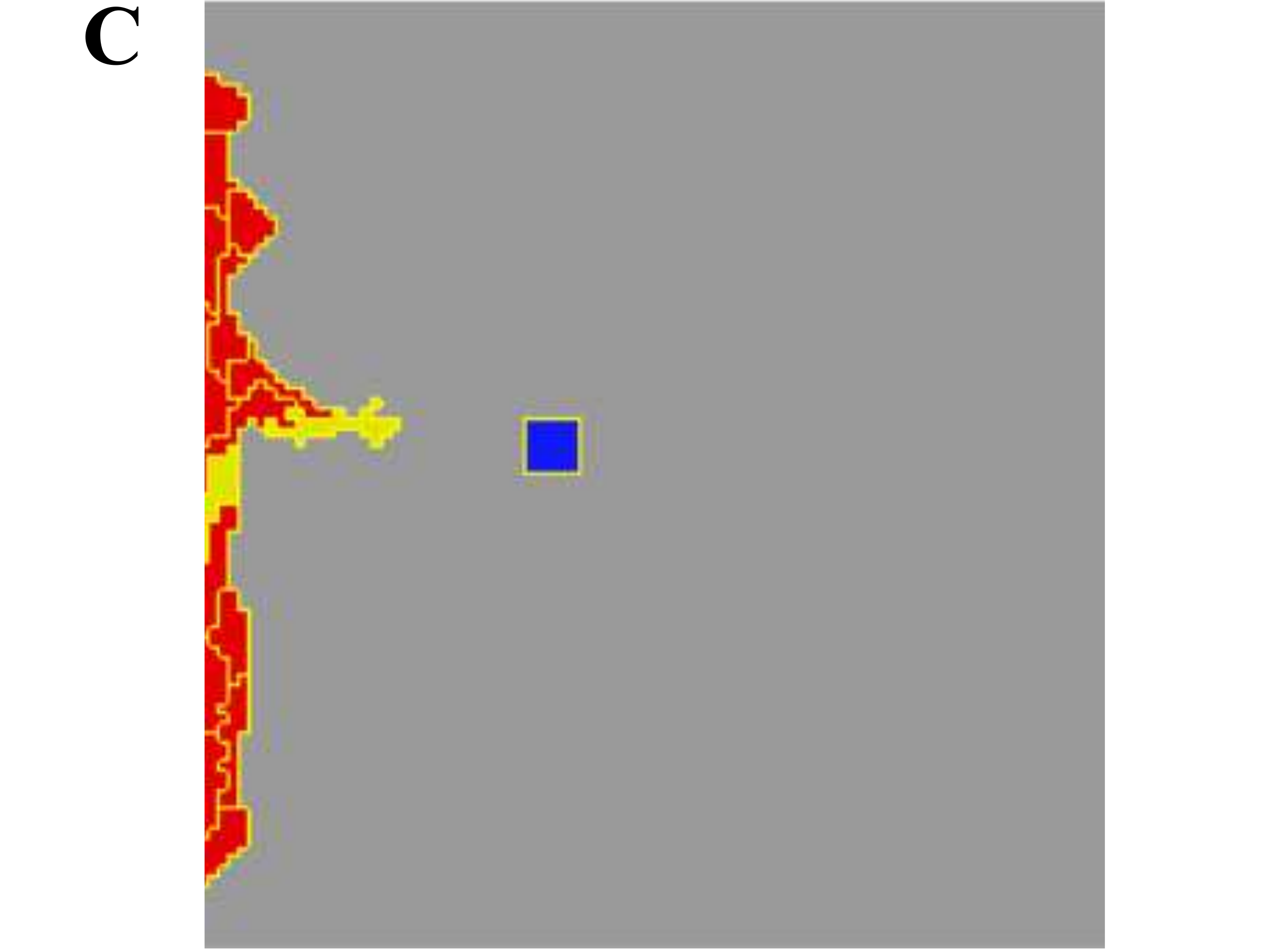} 
\label{fig:spr_morph_3}
\includegraphics[width=0.23\textwidth,angle=0]{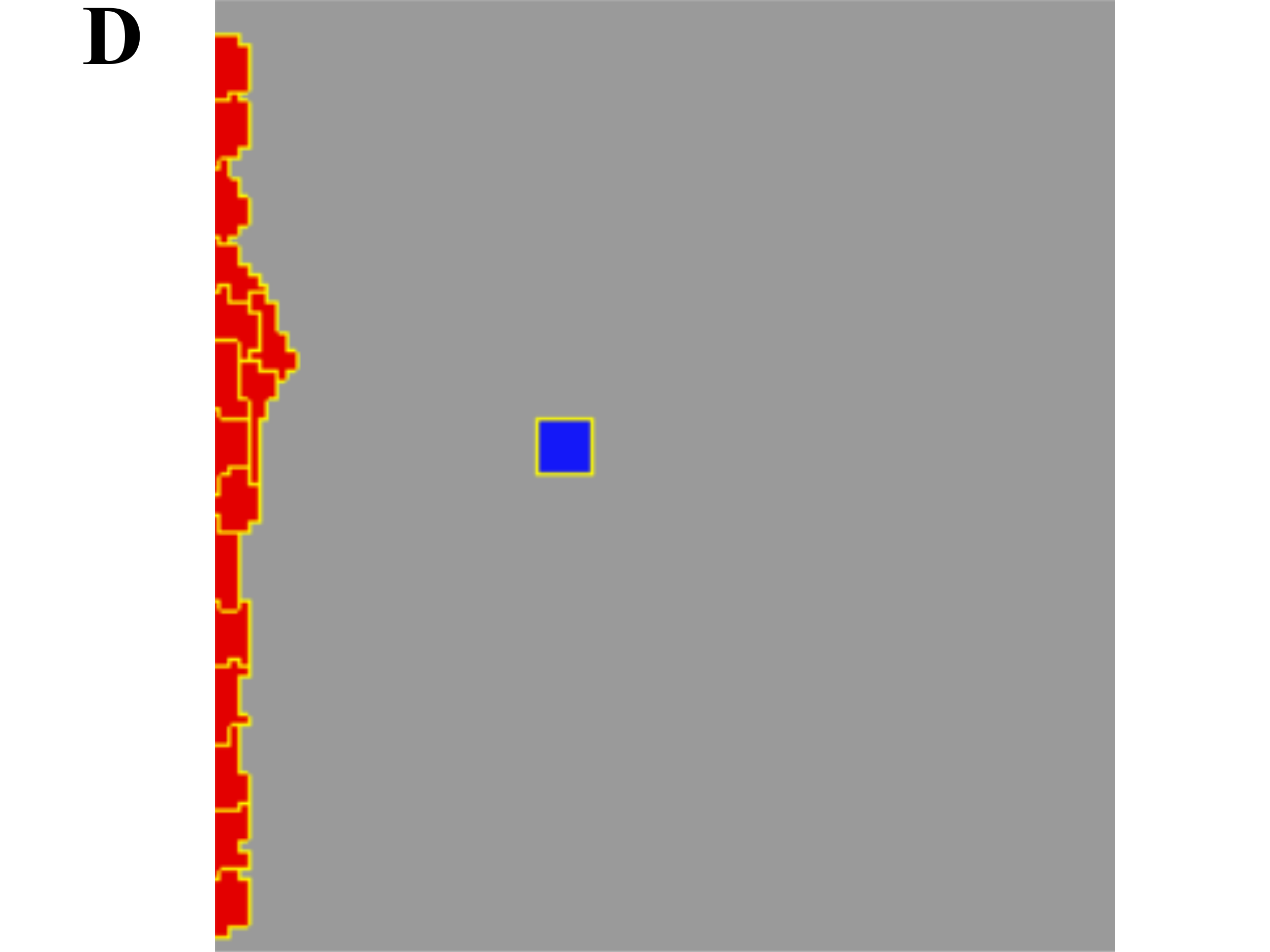} 
\label{fig:spr_morph_4}
\includegraphics[width=0.43\textwidth,angle=0]{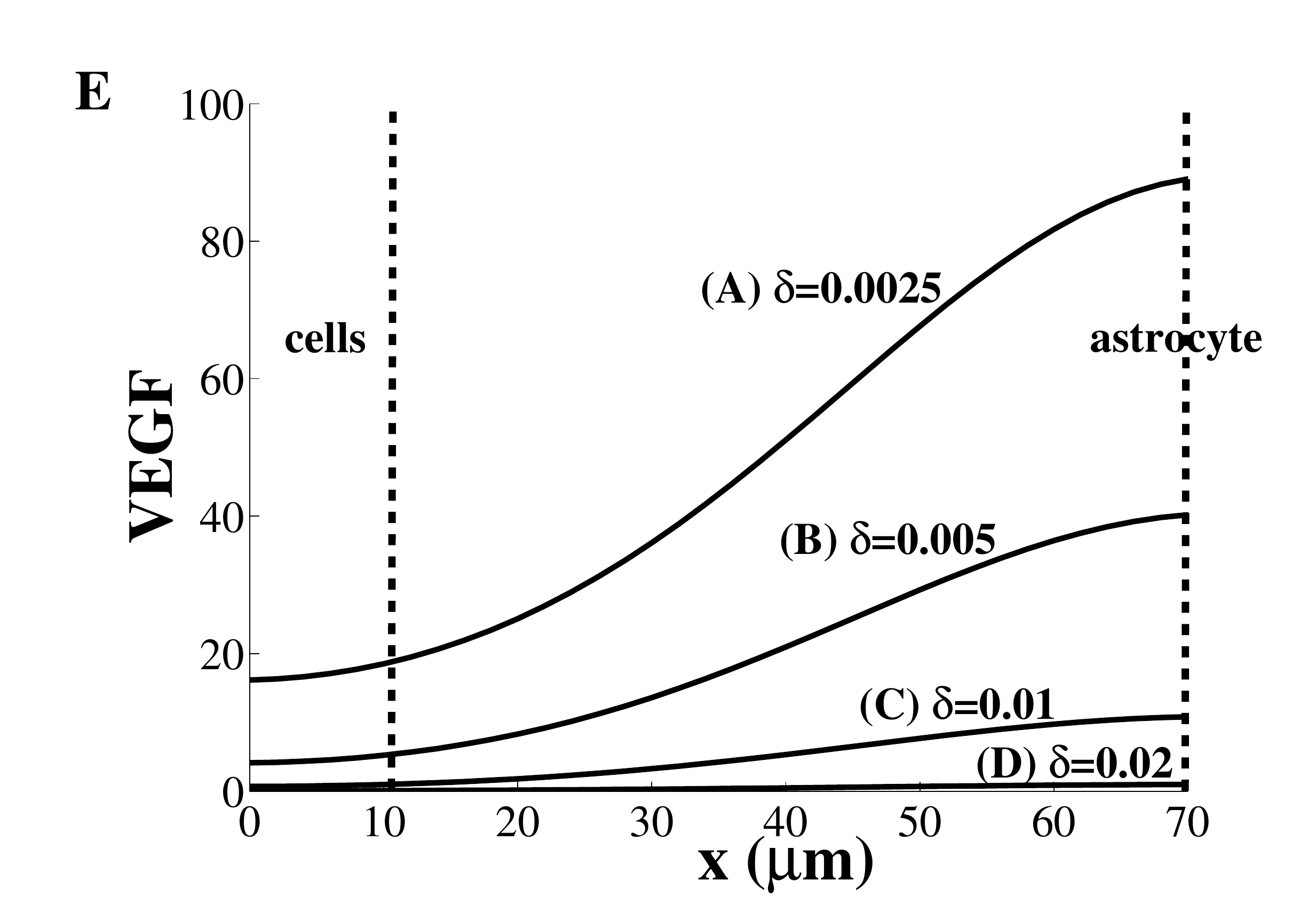} 
\label{fig:spr_morph_4}
\end{center}
\caption{\textbf{Variation of VEGF decay rate ($\delta$).} The markedly different capillary sprout morphologies (after $\sim$5 hrs) that result from small (A)-(B) $\delta=0.0025, 0.005$ respectively, intermediate (C) $\delta=0.01$, and large (D) $\delta=0.02$, VEGF decay rate $\delta$ (1/min). (e) Cross section of VEGF gradients for the four different cases of $\delta$. Cells of the parent blood vessel are located within 0-10 $\mu m$, and the astrocyte at $x=70 \mu m$. See Figure \ref{fig:setup} for the model setup. Key: stalk cells (red), tip cells (yellow), astrocyte (blue).}
\label{fig:decay_rate}
\end{figure}

Figure \ref{fig:decay_rate}(C) shows that intermediate decay rates imply directed sprouting.
However, in small decay rates (Figure \ref{fig:decay_rate}(A)-(B)) the sprout morphology dramatically changes.
The sprout splits off resulting in a mass of cells surrounding the fixed astrocyte, as a consequence of strong chemotaxis induced by steep VEGF gradients (Figure \ref{fig:decay_rate}(E)).
On the other hand, a large decay rate (Figure \ref{fig:decay_rate}(D)) may induce shallow VEGF gradients and low VEGF levels, which in turn do not allow for tip cell activation.

Increasing the VEGF decay rate could be equivalent to an anti-VEGF therapy characterized by the administration of e.g. a VEGF antibody a molecule which binds to VEGF with high affinity so that the free VEGF available to ECs is reduced \cite{Zheng2012}. Anti-VEGF therapy depends on the pericyte coverage of a sprout. In retina at day 10 all vessels are fully covered by pericytes 
and are resistant to anti-VEGF therapy \cite{Jo2006}.
Our model does not incorporate pericytes, which could be considered for future implementation. 

\vspace{0.5cm}

\FloatBarrier

\FloatBarrier
\textbf{Strength of chemotaxis:}\\
Chemotaxis promotes migration up the astrocyte-derived VEGFA gradients, and in this section, we investigate the effect of chemotactic strength, $\lambda_{chem}$, on the evolution and morphology of sprouting. The value of $\lambda_{chem}$ might depend on the amount of receptors for the chemoattractant expressed by the cells \cite{Shi1998}, and is hard to get values in measurable units. Therefore, a range of values was tested to find a close approximation to a growing vascular sprout. 

\begin{figure}[h!]
\begin{center}
\includegraphics[width=0.235\textwidth,angle=0]{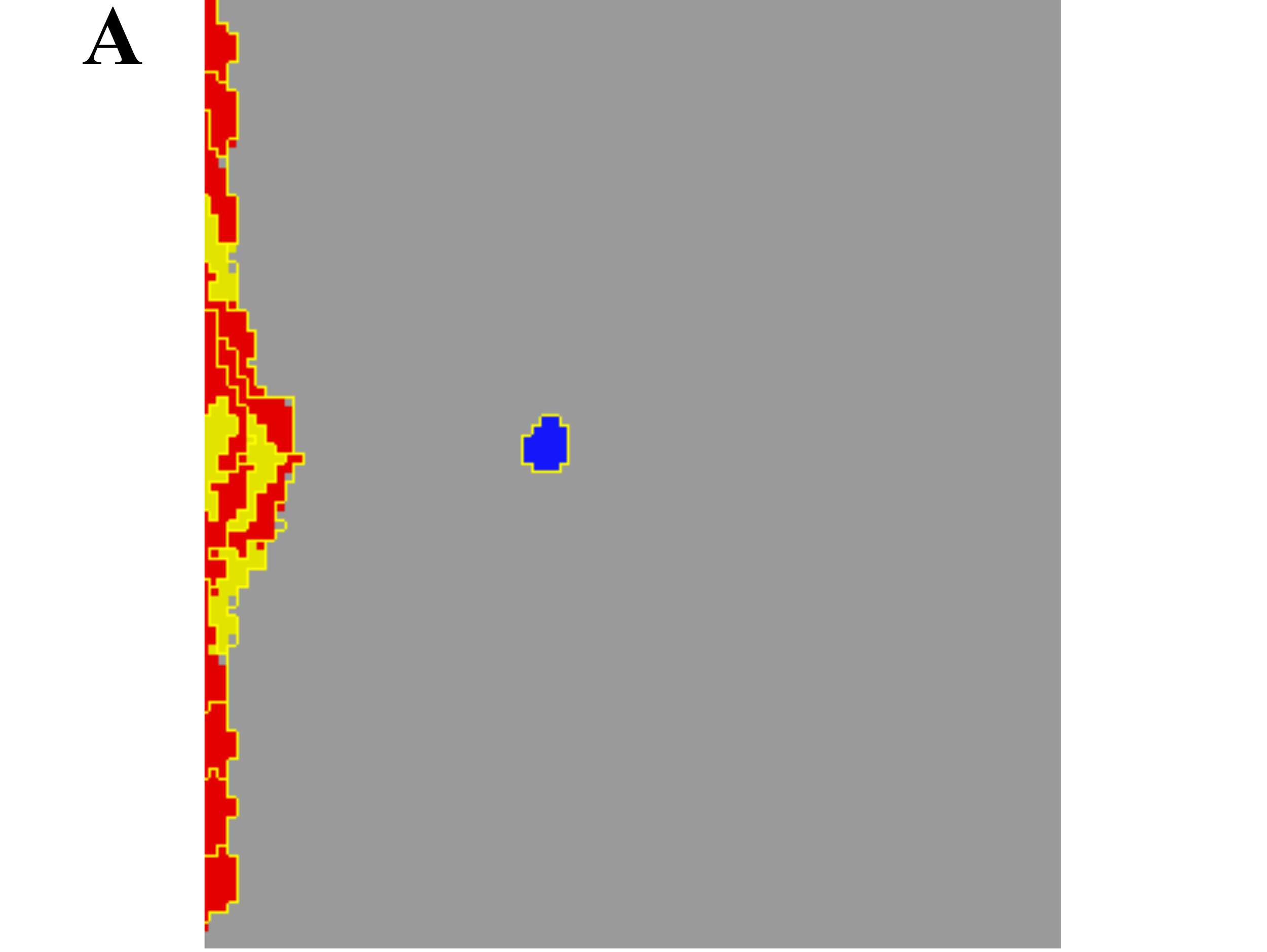}
\label{fig:spr_morph_1}
\includegraphics[width=0.235\textwidth,angle=0]{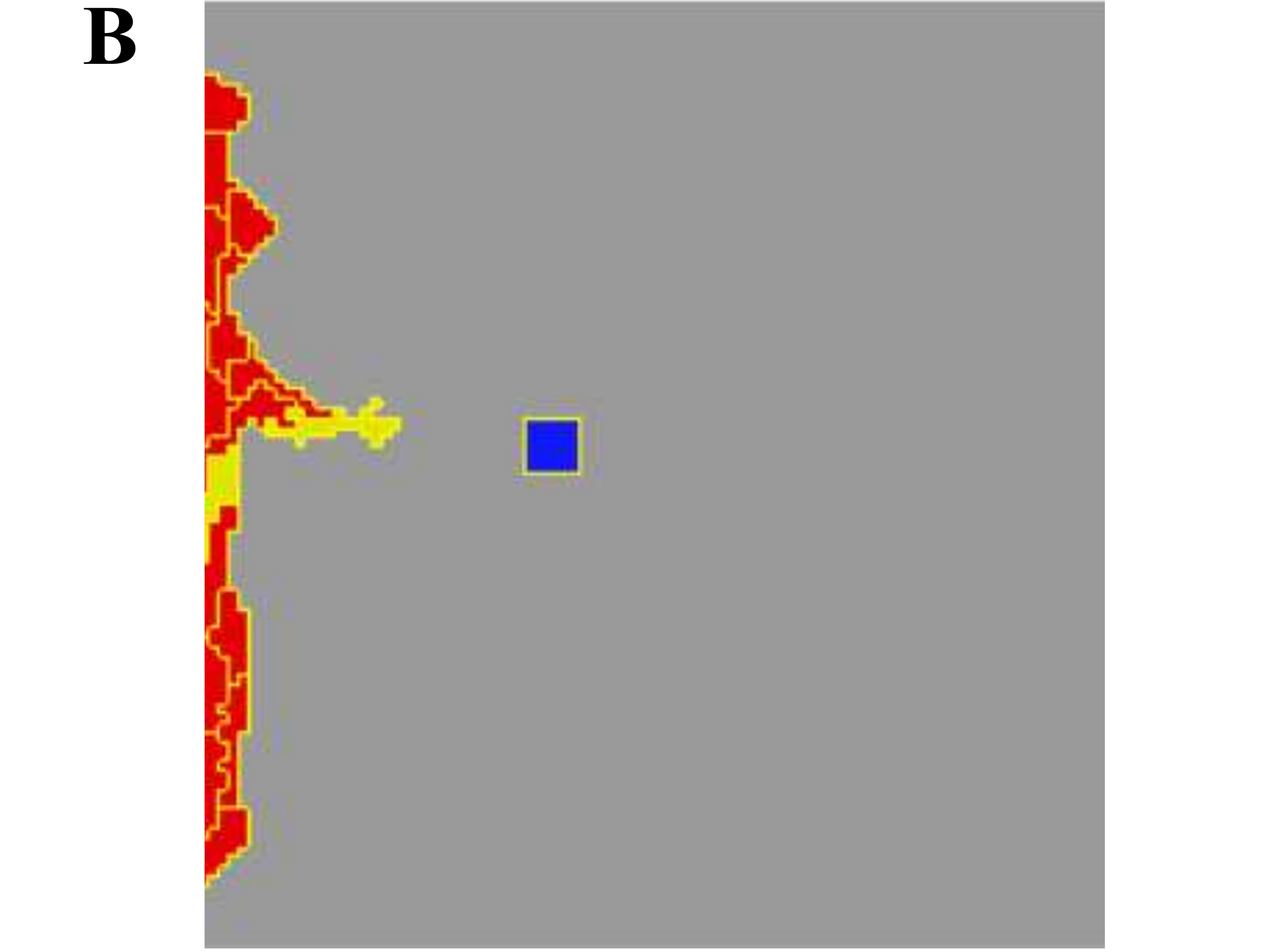} 
\label{fig:spr_morph_2}
\includegraphics[width=0.235\textwidth,angle=0]{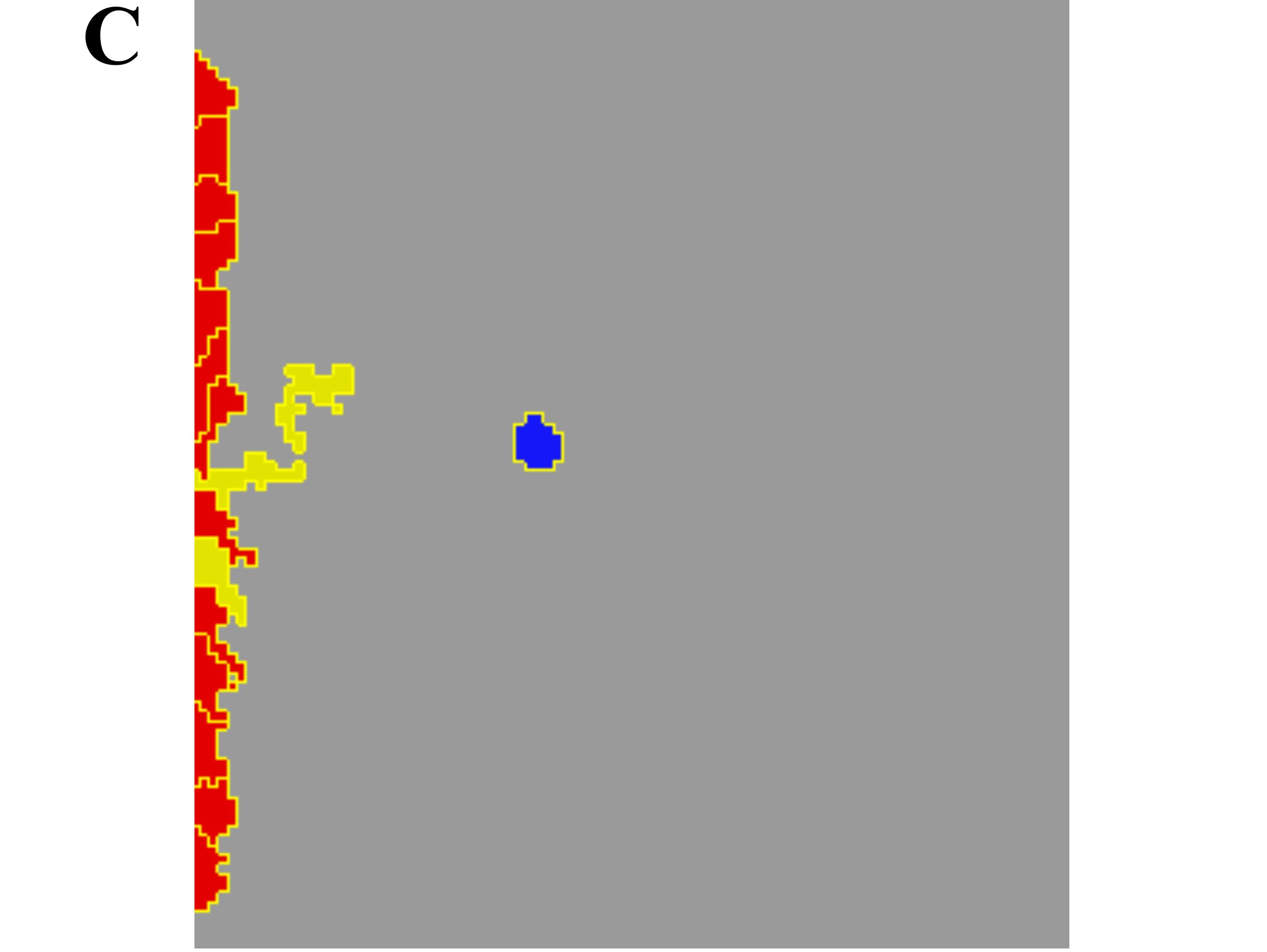}
\label{fig:spr_morph_3}
\includegraphics[width=0.235\textwidth,angle=0]{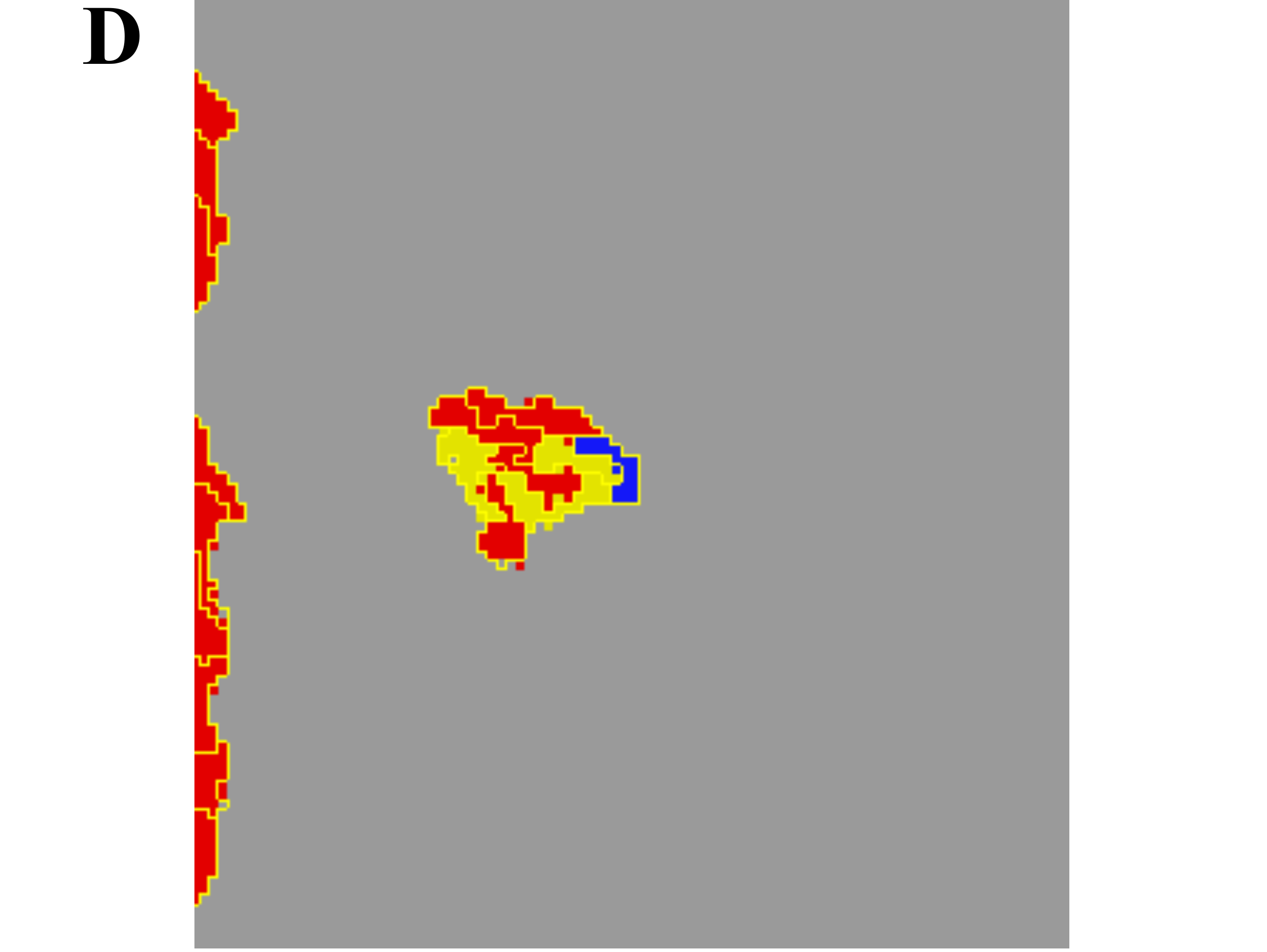}
\label{fig:spr_morph_4}
\end{center}
\caption{\textbf{Variation of the strength of chemotaxis ($\lambda_{chem}$).} Snapshots of growing sprouts after 5 hrs (except (C) 1 hr) by varying $\lambda_{chem}$ in three different cases, (A) low; $\lambda_{chem}=10$, (B) intermediate; $\lambda_{chem}=200$, and (C)-(D) high $\lambda_{chem}=500$.}
\label{fig:chemot}
\end{figure}

In Figure \ref{fig:chemot} we show morphological results of sprout formation from three examples: (A) low, (B) intermediate, and (C)-(D) high $\lambda_{chem}$. Figure \ref{fig:chemot}(A) shows that if strength of chemotaxis from tip cells is low that will eventually prevent sprouting (sprout will not reach the astrocyte). However, if it is large (Figure \ref{fig:chemot}(C)-(D)), the sprout splits off, which is comparable with low VEGF decay rate as it is shown in Figures \ref{fig:decay_rate}(A)-(B).
Strong chemotaxis can increase the extension speed and promote cell elongation, but the stronger the chemotaxis, the stronger the cell-cell adhesion is needed to keep cells in contact and avoid cell detachment.
Therefore, a better chemotactic response is achieved at intermediate $\lambda_{chem}$ (Figure \ref{fig:chemot}(B)), where there is a balance between cell-cell adhesion and chemotactic migration.

\vspace{0.5cm}

\FloatBarrier

\begin{figure}[h!]
\begin{center}
  \includegraphics[width=3.5in]{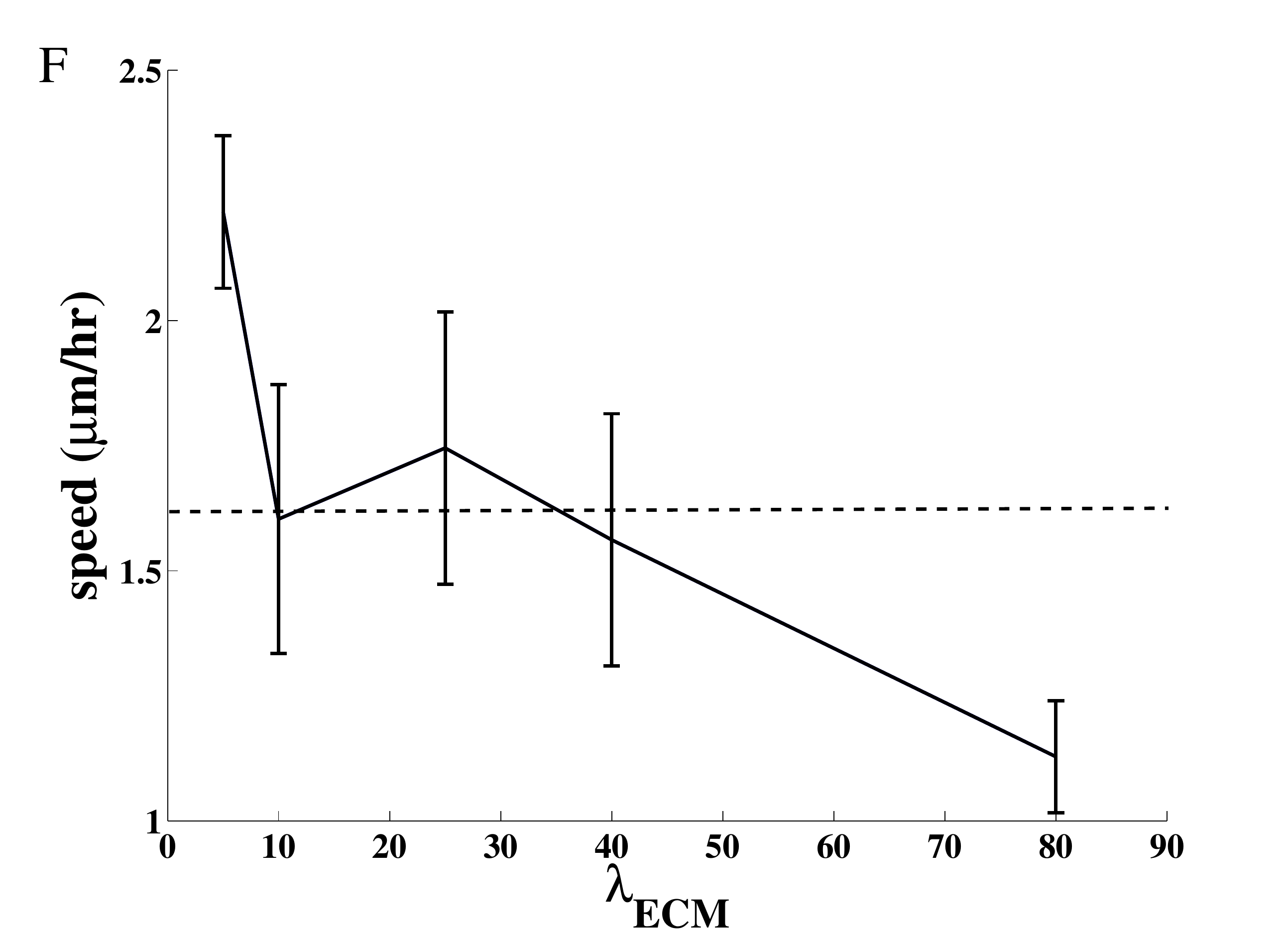} 
\end{center}
\caption{\textbf{Speed of sprout over $\lambda_{ECM}$.} We investigate how much the strength of preferential attachment to ECM, $\lambda_{ECM}$, can affect the speed of the sprout. It is shown that the higher the $\lambda_{ECM}$, the stronger the attachment to the ECM, which implies lower speed of the sprout. Error bars show the mean of 10 simulations $\pm$ S.E.M. The dashed line corresponds to the average extension speed ($\sim$3.5 $\mu m/hr$) from our experimental data at postnatal day 7.}\label{fig:speed_lambda_ECM}
\end{figure}

\FloatBarrier
\textbf{Strength of preferential attachment to ECM:}\\
Next we investigate the role of haptotaxis on the speed and morphology of the sprout formation by varying the strength of preferential attachment to ECM, $\lambda_{ECM}$ (see equation (\ref{eq:haptokinesis})).
Figure \ref{fig:speed_lambda_ECM} shows, as might be expected, that as $\lambda_{ECM}$ increases the speed decreases (e.g. $\sim$2 $\mu m/hr$ for $\lambda_{ECM}=100$) and the sprout is not able to reach the astrocyte since cells strongly adhere to the ECM.
However, intermediate values (e.g. $\lambda_{ECM}=60$; default value) give a good speed approximation to experimental data ($\sim$3.5 $\mu m/hr$).


\FloatBarrier

\FloatBarrier
\subsubsection*{Feedback between astrocyte and the new sprout}
\label{sec:modelvariants}
Empirical observations show that high concentrations VEGF is always located in the sprout front (see Figure \ref{fig:Luisa_macrophage}(B)), suggesting a reciprocal interaction between the sprout and the source cells \cite{West2005}. Two possible mechanisms are: 1) the VEGF source cells are moving away from the expanding sprout fronts, and 2) the VEGF source cells are spatially fixed, but only actively produce VEGF when interact with sprout tip. In both cases, it means a moving VEGF source.  We consider the case of VEGF source cell migration. 

As we described in Figure \ref{fig:incrementaloverview}, scenario 6 provides a close approximation to a growing vascular sprout.
However, astrocytes might not in fact be fixed in space, and therefore, we extend our model (in scenario 6) by allowing astrocyte motility to study the effect of a motile VEGF source. This is implemented by incorporating in the energy equation (\ref{eq:Hamiltonian}) the following constraint
\begin{equation}\label{eq:mac_motility}
H'_{motility} = r(cos\theta(t),sin\theta(t))\cdot(\vec{x} - \vec{x'}),
\end{equation}
with $r$ determining the cell speed, and $\theta$ the rotation angle.
Two important questions arise at this point: first, regarding the speed of the astrocyte, and second, the time point that the astrocyte should start moving. 
A reasonable answer to start with would be to allow the astrocyte to start moving from the beginning of each simulation.

\vspace{0.5cm}

\FloatBarrier
\textbf{Astrocyte moves from the beginning: sprout splitting}\\
We start by allowing the astrocyte to move from the beginning of each simulation, and with $\theta=0$ so that it moves preferentially parallel to the x-axis and to the right-hand side (away from the blood vessel). Since $r$ represents the speed, it would be appropriate to assess the sprout behavior in various values of $r$. In doing this, we found that for low/medium values (with speed below 3.5 $\mu m/hr$), the sprout can reach the astrocyte, and henceforth, the dynamics of the sprout have a strong effect on the motility and direction of the astrocyte because of cell-cell adhesion and chemotaxis. Reassuringly, when $r$ lies within low/medium ranges the sprout forms up to a certain time point. However, the sprout eventually splits, presumably at the point where the VEGF gradients become less steep. We observed that after 42 hrs the detached section of the sprout continues to `push' (chemotacting towards) the astrocyte.
On the other hand, if $r$ is large (and speed becomes greater than 3.5 $\mu m/hr$), the sprout is no longer able to reach the astrocyte, and eventually all the cells in the blood vessel become or remain as stalk. Therefore, in this case we get no sprout formation (morphological results not shown).
%

\vspace{0.5cm}

\begin{figure}[h!]
\begin{center}
\includegraphics[width=0.235\textwidth,angle=0]{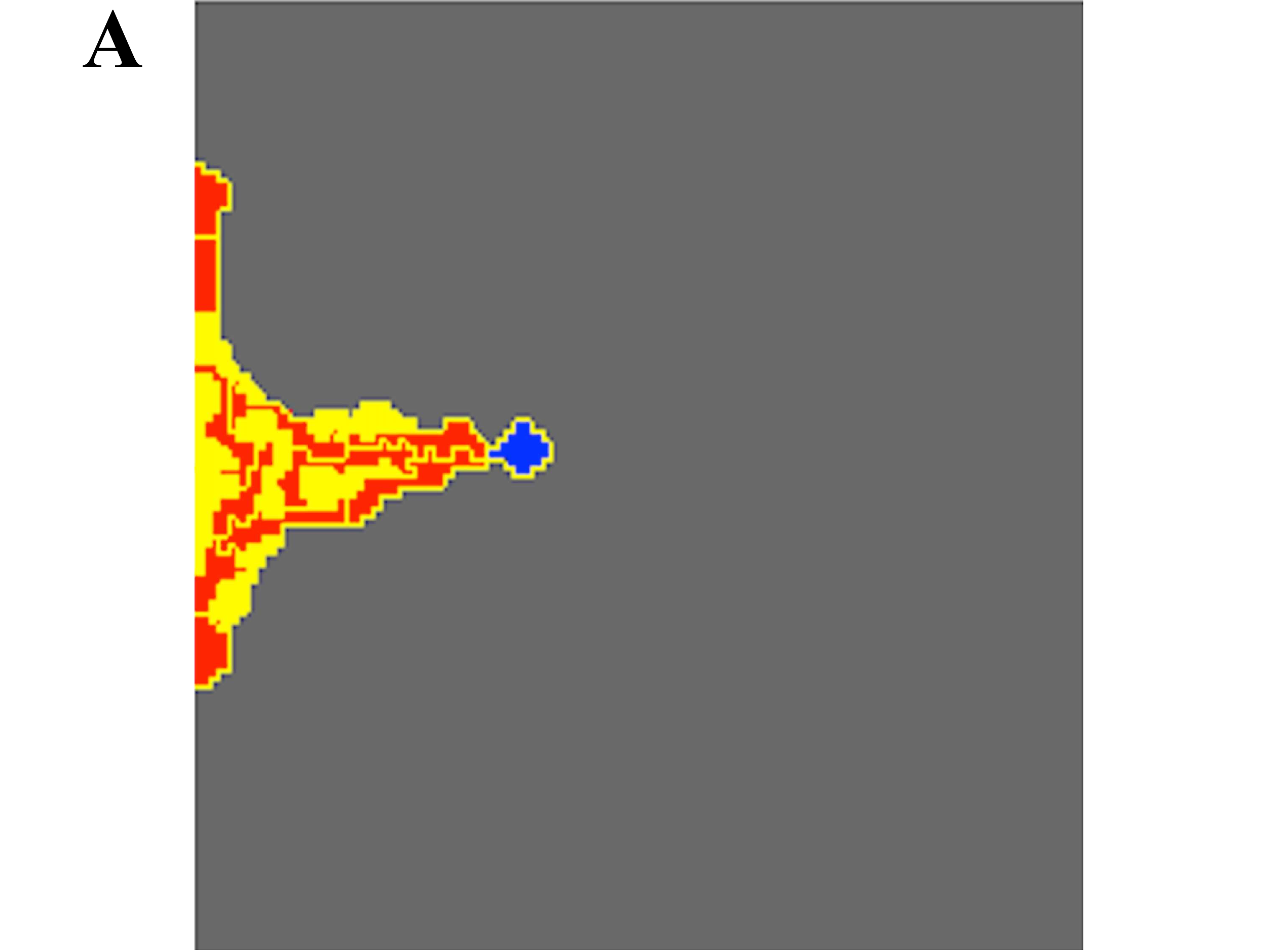}
\label{fig:spr_morph_1}
\includegraphics[width=0.235\textwidth,angle=0]{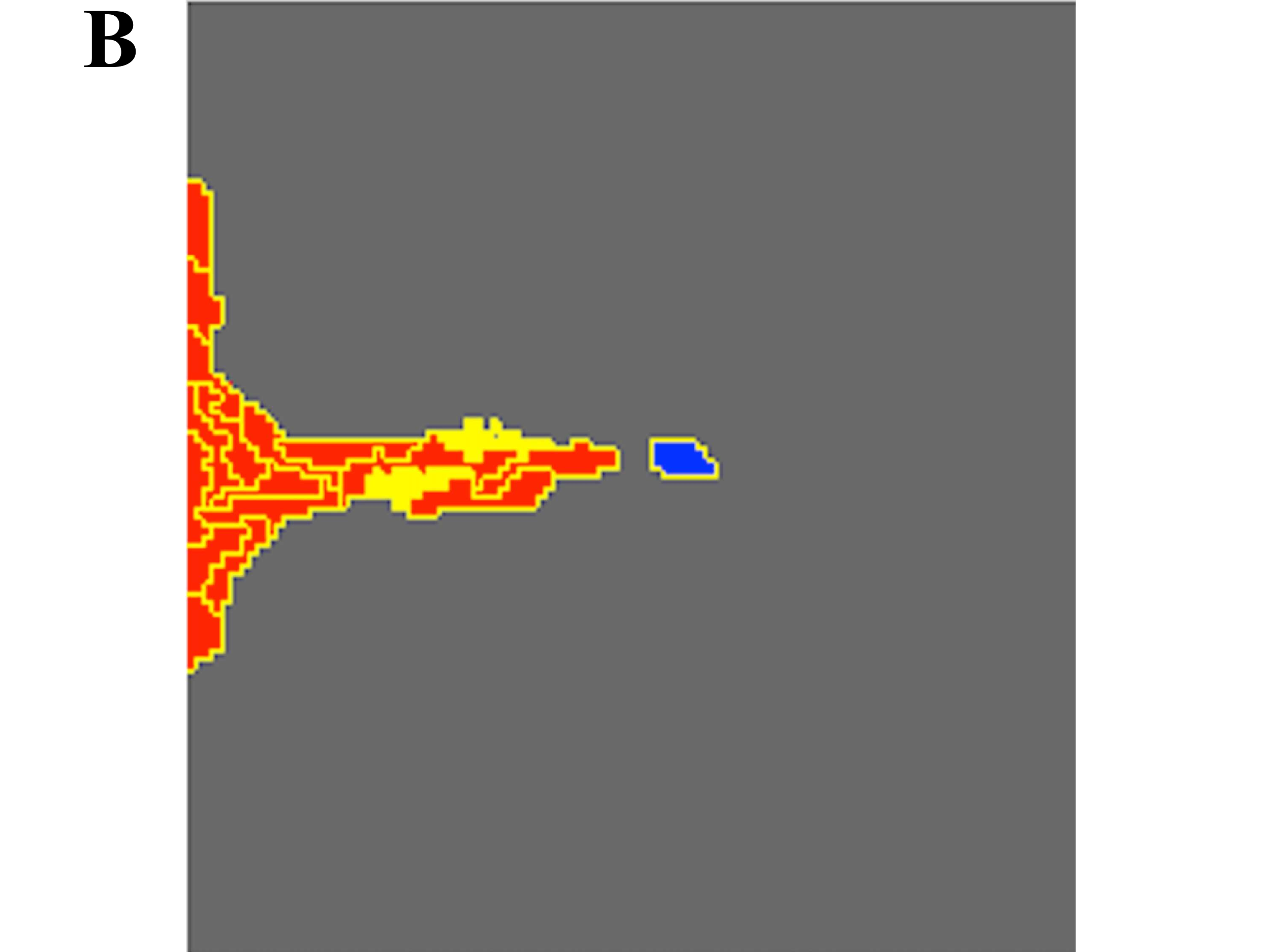}
\label{fig:spr_morph_2}
\includegraphics[width=0.235\textwidth,angle=0]{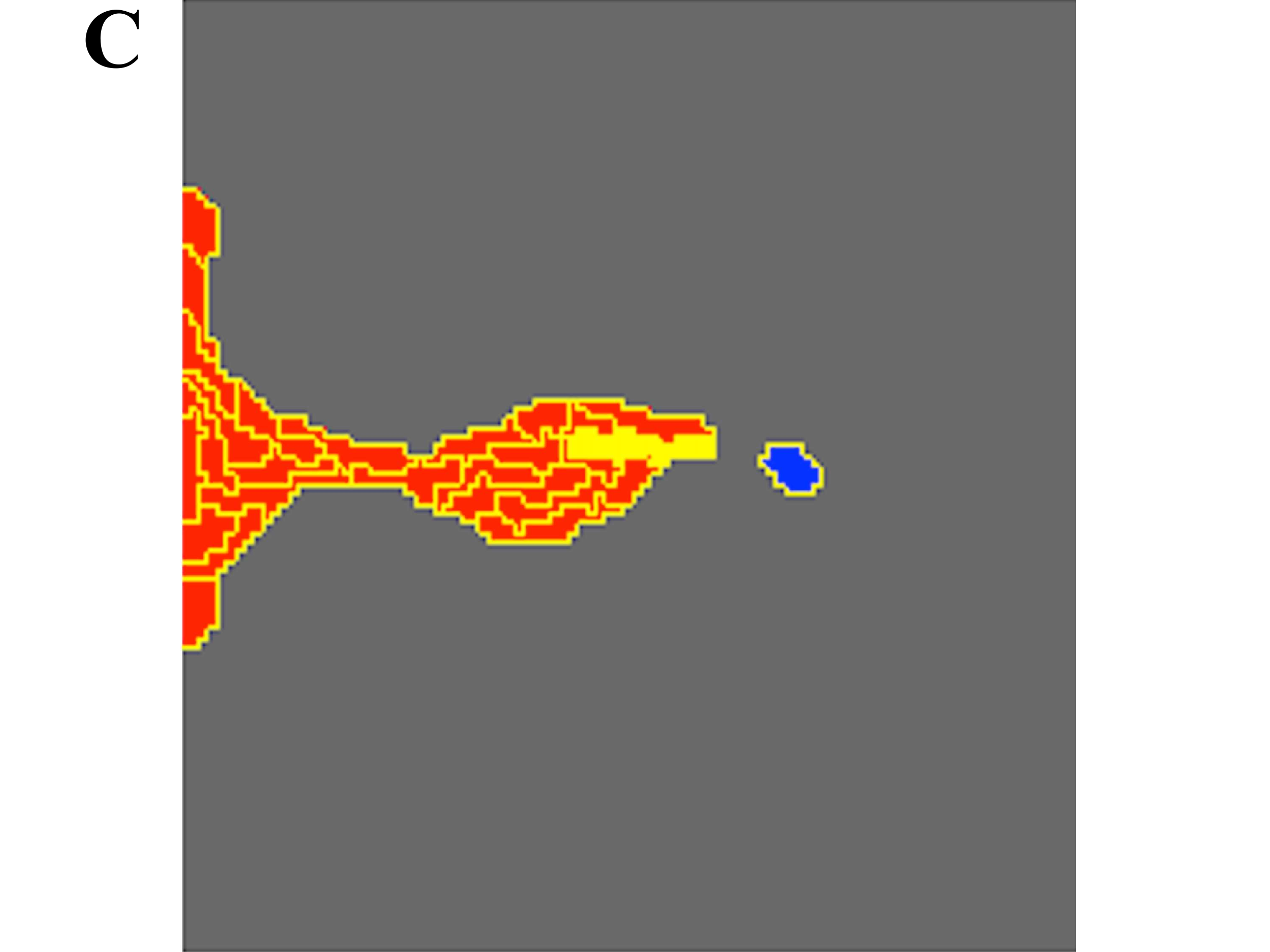}
\label{fig:spr_morph_3}
\includegraphics[width=0.235\textwidth,angle=0]{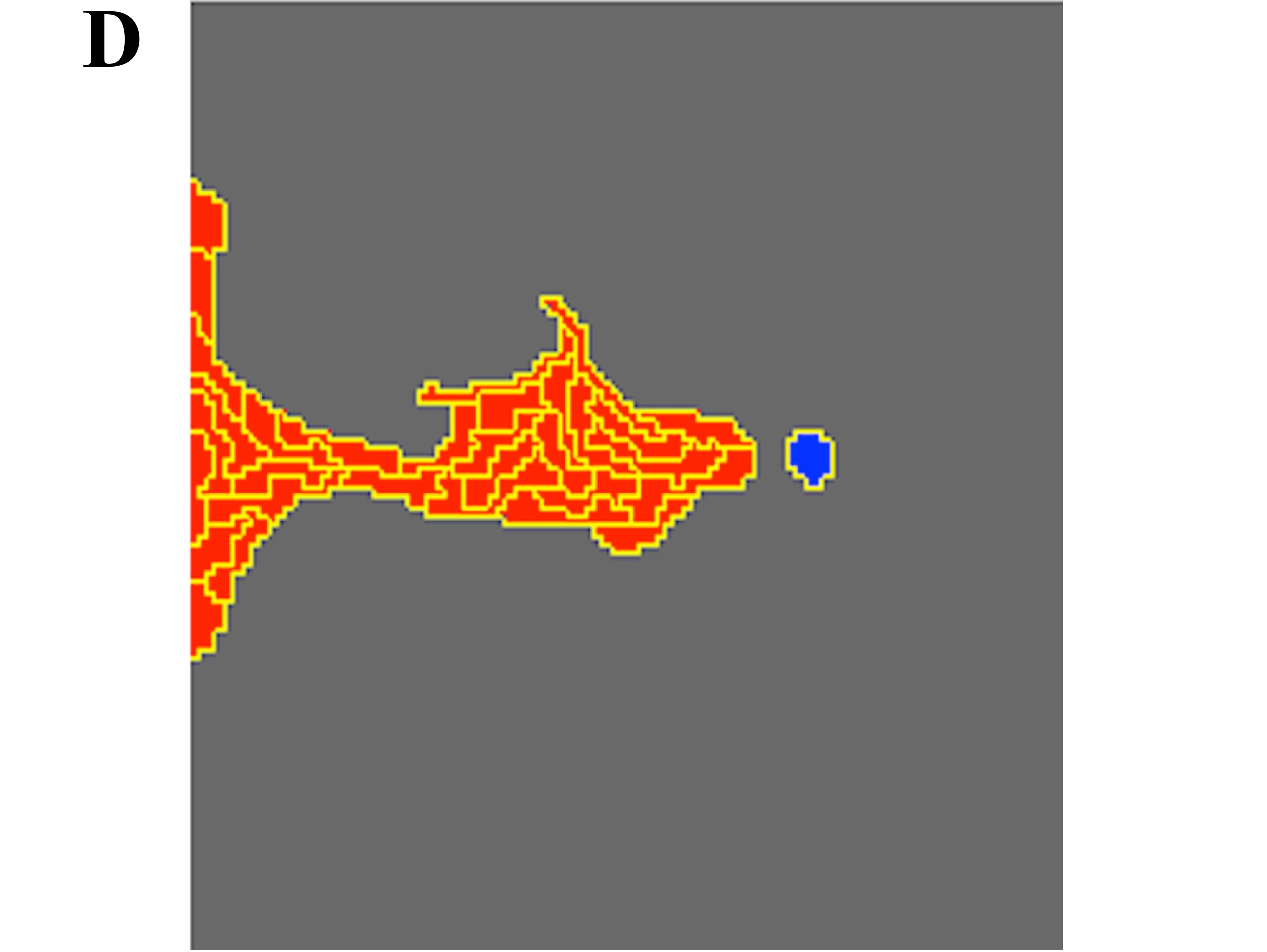}
\label{fig:spr_morph_4}
\end{center}
\caption{\textbf{Sprout evolution with a moving astrocyte.} Representative snapshots (A) 6 hrs, (B) 24 hrs, (C) 48 hrs, (D) 72 hrs show that by adding motility to the astrocyte in scenario 6 can result in a long sprout formation without sprout splitting as is the case with a fixed astrocyte. Astrocyte motility is described as in equation (\ref{eq:mac_motility}), and the astrocyte moves only when in contact with the an endothelial stalk or tip cell. Key: stalk cells (red), tip cells (yellow), astrocyte (blue).}\label{fig:movmacrophage}
\end{figure}

\FloatBarrier
\textbf{Astrocyte moves only when in contact with the sprout: long sprout formation}\\
The above results suggest that some form of coupling between sprout and astrocyte movement might be important. Thus we incorporate a mechanism in which the astrocyte can move only when it comes in contact with an endothelial stalk or tip cell, and we find that this does result in a straight sprout without cell detachment (see Figure \ref{fig:movmacrophage} for an example).
Although the mechanism is not well understood, this seems to be consistent with a concept suggesting the existence of a feedback between astrocyte and the ECs of the new sprout \cite{West2005}.
%

\FloatBarrier

\FloatBarrier
\section*{Discussion}
In this paper, we have developed a 2-D multiscale Cellular Potts Model (CPM) with the aim to understand the dynamic interaction between stalk and tip cells (two endothelial cell (EC) phenotypes) during sprout formation integrated via the VEGFA-Delta-Notch signaling pathway.

Sprouting angiogenesis requires activation of normally quiescent ECs in pre-existing blood vessels, breakdown of existing basement membranes, migration of activated cells led by one or more endothelial tip cells and proliferation of a subset of activated ECs (stalk cells).
The complex biological processes leading to sprout formation are a consequence of cell-level decisions that are based on global signals (e.g. VEGFA signaling) and juxtacrine communication (e.g. Notch-Delta signaling). In particular, extracellular VEGFA activates the intracellular Notch-Delta pathway in ECs which result in endothelial (stalk, tip) differentiation.
\emph{In vivo} retinal angiogenesis experiments reveal the crucial role that astrocytes seem to have in polarizing new sprouts by secreting VEGFA, highlighting the growing evidence for a link between astrocytes and angiogenesis \cite{Scott2010}.

In our model we assessed the effects of six different scenarios regarding homogeneous and heterogeneous VEGFA and extracellular matrix (ECM) profiles on sprout morphology. In particular, two with homogeneous/uniform VEGFA (no gradients), two with static VEGFA gradients, and two with VEGFA gradients emerged from a fixed astrocyte. In each pair of scenarios we alternated between uniform and non-uniform (represented as a network of static fibers) ECM.
Experimental studies have revealed a dynamic shuffling of tip and stalk cells at the leading front of growing sprouts challenging the idea of stable tip and stalk cell selection \cite{Geudens2011}.
This dynamic behavior was incorporated in our model (compared to other mathematical models using fixed cell positions) in the following way.
A stalk cell may become a tip cell if its Delta level, which is upregulated from the astrocyte-derived VEGFA, exceeds a threshold value. In turn, a fine-tuned feedback loop between VEGFA and Notch-Delta signaling establishes a `salt-pepper' distribution (checkerboard pattern) of stalk and tip cells within the activated endothelium.
Cells continually respond via their VEGFA-Delta-Notch signaling loop when they meet new neighbors,
and the lateral-inhibition effect via the Notch signaling defines the (interchangeable) fate of the (stalk and tip) cells in the new sprout. Stalk cells adjacent to tip cells may proliferate and elongate to support sprout elongation. The proliferation of stalk cells was based on the cell cycle time estimated from available retinal experimental data. Tip cells are allowed to elongate, but not to proliferate \cite{Gerhardt2003}. The `head' tip cell leads the sprout forward distinguishing it from other tip cells in the sprout having the maximum VEGFA level compared to any other cell.
As the stalk and tip cells migrate through the ECM following up the chemotactic (moving up VEGFA gradients) and haptotactic cues (moving preferably along ECM fibers if in a scenario with non-uniform ECM), they define the morphology of the outgrowing sprout, with tip cells leading the sprout polarization.
Tip cells contact other tip cells (anastomosis) to extend the existing vascular network, a result which is also dynamically captured in our model.
%

By considering the sprout morphological dynamics from all six scenarios, we may conclude that narrow sprout formation can be closely approximated under scenario 6 with heterogeneous VEGFA (with fixed astrocyte) and heterogeneous (non-uniform) ECM in $\sim$18 hrs. However, at later time points cells start surrounding the fixed astrocyte without being able to produce a longer sprout formation (see Figure \ref{fig:incrementaloverview}). In fact astrocytes might not be fixed \cite{Zhang1999} and, therefore, we extended our model by incorporating astrocyte motility (moving away from the blood vessel) in order to test its effect on sprout formation. Interesting question which arise at this point concern whether the movement of the sprout and the astrocyte are coupled or not. At first, we simply allowed the astrocyte to start moving from the beginning of a simulation, and we assessed various speeds of the astrocyte.
We found that low or medium astrocyte speeds could result in sprout formation of a limited length because of sprout splitting. On the other hand, high speed resulted in no sprouting because astrocyte moves too fast and VEGFA gradients do not allow for tip cell activation. Therefore, all of the cells in the blood vessel remained as stalks. However, by coupling astrocyte movement to EC movement, simulations show the emergence of a straight and long sprout along the whole numerical domain without splitting (see Figure \ref{fig:movmacrophage}).
Although further experimental work needs to be done in the future, the latter result seems to be consistent with the existence of an endothelial (sprout)-astrocyte feedback \cite{West2005}.

%
Sensitivity analysis was performed for key parameters, such as the VEGFA decay rate ($\delta$; Figure \ref{fig:decay_rate}), the strength of chemotaxis ($\lambda_{chem}$; Figure \ref{fig:chemot}), and the strength of the attachment to the ECM ($\lambda_{ECM}$; Figure \ref{fig:speed_lambda_ECM}).
In particular, in basal decay rates there is directed sprouting. However, in small $\delta$ the sprout eventually splits, whereas in large $\delta$ there is no tip cell activation and, therefore, no sprout formation.
Regarding $\lambda_{chem}$, if it is small, the sprout is not able to reach the astrocyte, whereas in large $\lambda_{chem}$ the sprout splits. It was also shown that the value of $\lambda_{ECM}$ affects the speed of the sprout. As might be expected, if $\lambda_{ECM}$ is large the speed of the sprout falls below the available experimental  measurements since cells strongly adhere to the ECM.

We also performed VEGFA-Notch-Delta knockout \emph{in silico} experiments in uniform VEGFA environment as an attempt to test that as an anti-angiogenic treatment regimen.
Recall that extracellular VEGFA upregulates Delta levels in each EC. Interestingly, results show that decreasing the sensitivity of Delta to VEGFA (parameter $\alpha$ in equation (\ref{eq:ODEmodel})), sprout formation is not possible.
Our intention is to delve more into the therapeutic aspect (e.g. blocking sprout formation) in the future. Pericytes, which surround the ECs, are now coming into focus as important regulators of angiogenesis and blood vessel function, with the capacity to be resistant to anti-VEGF therapy \cite{Jo2006}. Therefore, our model could be extended by incorporating pericytes in order to assess its role on therapeutic regimes.

In summary, we have developed a multiscale model that incorporates intra-, inter-, and extra-cellular levels for studying sprout evolution in angiogenesis. Several biological hypotheses have been considered regarding the VEGFA source (e.g. astrocyte), Notch-Delta pathway activation from VEGFA, cell chemotaxis, cell-cell adhesion, as well as cell-ECM contact guidance. Simulation results have been successful in reproducing sprouting morphogenesis, with the extension speed of the sprout being in agreement with experimental data in 7 days postnatal mouse retina. Interestingly, the results with moving astrocyte (VEGFA source) suggest that coupled movement of endothelial cell and astrocyte ensures long and contiguous sprout formation.
This observation is consistent with the proposal, suggesting the existence of interactions between sprout and astrocytes \cite{West2005}. Our model framework will allow detailed, systematic investigation on these interactions, which will be a topic of our future study.

\FloatBarrier
\section*{Supporting Information}

\FloatBarrier
\subsection*{Experimental methods}
\label{sec:homogeneous}
%
Eyes were isolated and stained as reported in \cite{HofmannArispe2007}. Retinas were taken from P5 pups. Retinas were stained using antibodies against PECAM (MEC 13.3, BD Pharmingen, CA) and fibronectin (9661S, Millipore, Billerica, MA). A Zeiss LSM 510 META confocal microscope was used to image the retinas and the Zen software was used for acquisition (Carl Zeiss Microscopy, Cambridge, UK).

\FloatBarrier
\subsection*{Analysis of perturbations of the homogeneous steady state}
\label{sec:homogeneous}
In the following we present mathematical analysis on perturbations of the homogeneous steady state of equations (\ref{eq:ODEmodel}), and parameter ranges in which the `salt-pepper' pattern is maintained.

The non-dimensional model as given in equations (\ref{eq:ODEmodel}) can be rewritten in the following form
\begin{eqnarray}
\begin{array}{lll}
\vspace{0.3cm}
\dfrac{d D_j}{dt}= v\left( \alpha h(V)g(N_j)  - D_j \right) \;, \\
\vspace{0.3cm}
\dfrac{d N_j}{dt}= f(\bar{D}_j) - N_j\;,\\
\end{array}
\label{eq:ODEmodel_new}
\end{eqnarray}
where $h(V) = \dfrac{VEGF}{VEGF_h + VEGF}$, $g(N_j)=\dfrac{1}{1 + bN^{2}_j}$,
$f(\bar{D_j}) = \dfrac{\bar{D}_j^2}{a + \bar{D}_j^2}$, and $\bar{D}_j= \sum_{i} \dfrac{D_i P_{ij}}{P_j}$ with $P_j, P_{ij}$ being the perimeter of cell $j$, and the common area between neighbor cells $i$ and $j$, respectively. Note that $f,g$ are continuously differentiable, with $f$ monotonic increasing and $g$ monotonic decreasing. Under these conditions there exists exactly one homogeneous steady state (HSS), $(D^*_j,N^*_j)=(\alpha h(V)g(N^*_j),f(\bar{D^*_j}))=\left(\alpha h(V)g(N^*_j),f\left(\dfrac{D^*_j}{\sigma}\right)\right)$, where $\sigma=1, \textrm{if lattice sites are squares/hexagons};
\sigma=2, \textrm{if\:strings}$.
The HSS is defined as the steady state in which all cells have identical levels of Delta and Notch, and we wish to determine the patterns that emerge from perturbations about this steady state.
Therefore, we assume that $D_j = D^*_j + \tilde{D_j}$, $N_j = N^*_j + \tilde{N_j}$ for $\tilde{D_j},\tilde{N_j} \ll 1$, and by also making the following ansatz
\begin{equation}\label{eq:ansatz1}
\tilde{D}(\textbf{x},t) = \hat{D}(t)\exp^{ik\textbf{x}},
\end{equation}
which allows us to remove the spatial dependency $\textbf{x}$ of $D$,
we get
\begin{equation}\label{eq:ansatz2}
\bar{\tilde{D}} = \tilde{D}\dfrac{K}{\sigma},
\end{equation}
where $K(k)$, the `nearest neighbor contribution', is defined as
\begin{equation}
K(k) =
\left\{
\begin{array}{ll}
\vspace{0.3cm}
\:\cos(k)                                    & (\textrm{strings}), K\in[-1,1],\\
\vspace{0.3cm}
\dfrac{\cos(k_1) + \cos(k_2)}{2}            & (\textrm{squares}), K\in[-1,1] ,\\
\dfrac{\cos(k_1) + \cos(k_2) + \cos(k_1 + k_2)}{3} & (\textrm{hexagons}), K\in[-1/2,1] ,
\end{array}
\right.
\label{eq:K}
\end{equation}
with $k$ being the wavenumber (or wave-vector, $k=(k_1,k_2)$ in two space dimensions).

The system (\ref{eq:ODEmodel_new}) can be linearized about the HSS to give
\begin{eqnarray}
\begin{array}{lll}
\vspace{0.3cm}
\dfrac{d \tilde{D}}{dt}= v\left( \alpha h(V)G\tilde{N}  - \tilde{D} \right) \;, \\
\vspace{0.3cm}
\dfrac{d \tilde{N}}{dt}= F\tilde{D}\dfrac{K}{\sigma} - \tilde{N}\;,\\
\end{array}
\label{eq:ODEmodel_linearised}
\end{eqnarray}
where,
\begin{eqnarray}
\begin{array}{lll}
\vspace{0.3cm}
F=f'(\frac{D^*}{\sigma})=\frac{2a\frac{D^*}{\sigma}}{(a + (\frac{D^*}{\sigma})^2)^2}>0 \;, \\
\vspace{0.3cm}
G=g'(N^*)=-\frac{2bN^*}{(1 + bN^{*2})^2}<0 \;,\\
\end{array}
\label{eq:ODEmodel_FG}
\end{eqnarray}
and $'$ denotes differentiation. The linearized system (\ref{eq:ODEmodel_linearised}) gives the following Jacobian matrix
\begin{equation}
J(K) =
\left(
\begin{array}{ll}
-v        & v\alpha h(V)G \\
F\dfrac{K}{\sigma}  & -1
\end{array}
\right).
\label{eq:Jacobian}
\end{equation}
The trace, $tr(J(K))=-v-1$, is always negative since $v>0$, and
\begin{equation}\label{eq:det}
det(J(K))=v\left(1 - \alpha h(V)FG\dfrac{K}{\sigma} \right) = v\left(1 - \Phi\right).
\end{equation}
From stability analysis we know that if $tr(J)<0$ and $det(J)>0$ then we have a stable HSS.
As a model for lateral-inhibition, we are most interested in the case where the HSS is
stable to homogeneous ($k=0$), and unstable to heterogeneous ($k\neq0$) perturbations.
That is when $det(J(1))>0$, and $det(J(K))<0$ for some $K \in [-1,1)$ or $[-1/2,1)$. 

\begin{figure}[h!]
\begin{center}
  \includegraphics[width=2.8in]{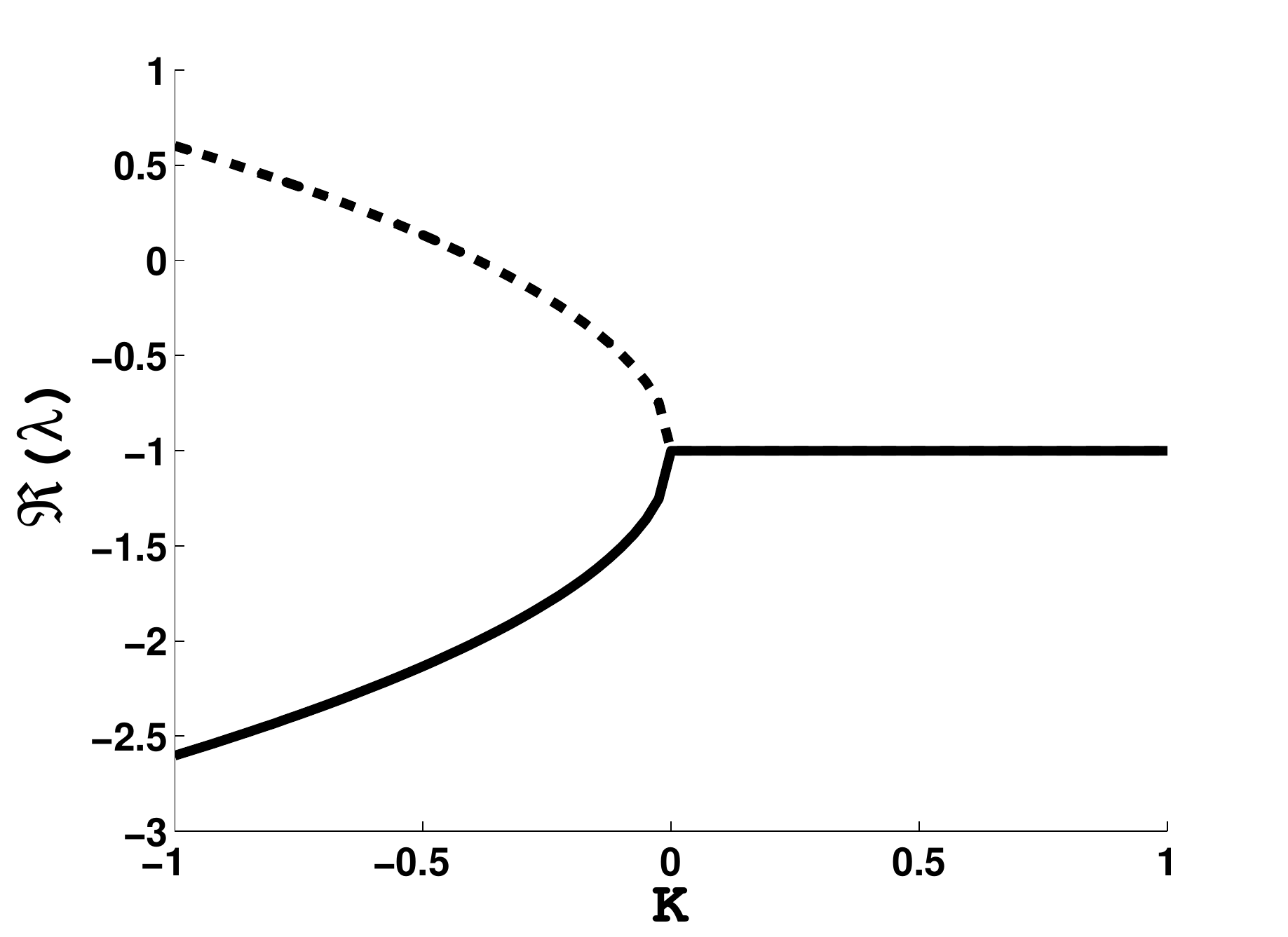}
\end{center}
\caption{\textbf{Real part of the eigenvalues over K.} The $\Re (\lambda_{1,2})$ of the eigenvalues $\lambda_{1,2}$ (dashed and solid curves), evaluated from the Jacobian as in (\ref{eq:Jacobian}), is maximal when $K=-1$ (for strings/squares as in (\ref{eq:K})). The parameter values used in the model (\ref{eq:ODEmodel_new}): $a=0.01$, $b=100$, $v=1$, $VEGF_h=1$, $VEGF=1$, $\alpha=1$.}\label{fig:eigenvalue}
\end{figure}

Recall that $G$ is always negative because of lateral-inhibition and, therefore, we exclude the cases when $K>0$ (for which $det(J)>0$). Clearly $det(J(1))>0$ since $F>0$ and $G<0$, so we only need to check the sign of $det(J(K))$.
Note that the real part of the eigenvalues of $J(K)$ is maximal for the smallest possible value of $K$ ($K=-1$ for strings/squares, and $K=-1/2$ for hexagons) - see Figure \ref{fig:eigenvalue}. Therefore a patterning instability requires
\begin{eqnarray}
\begin{array}{lll}
\vspace{0.3cm}
\textrm{strings:}\:\:\:\:\:\:\:\:\:  det(J(-1))<0 \Leftrightarrow \Phi=\alpha h(V)FG < -2,\\
\vspace{0.3cm}
\textrm{squares:}\:\:\:\:\:\:\:  det(J(-1))<0 \Leftrightarrow \Phi=\alpha h(V)FG < -1,\\
\vspace{0.3cm}
\textrm{hexagons:} \:\:\:  det(J(-1/2))<0 \Leftrightarrow \Phi=\alpha h(V)FG < -2.\\
\end{array}
\label{eq:FG_Collier}
\end{eqnarray}

\FloatBarrier

\begin{figure}[h!]
\begin{center}
\includegraphics[width=0.4\textwidth,angle=0]{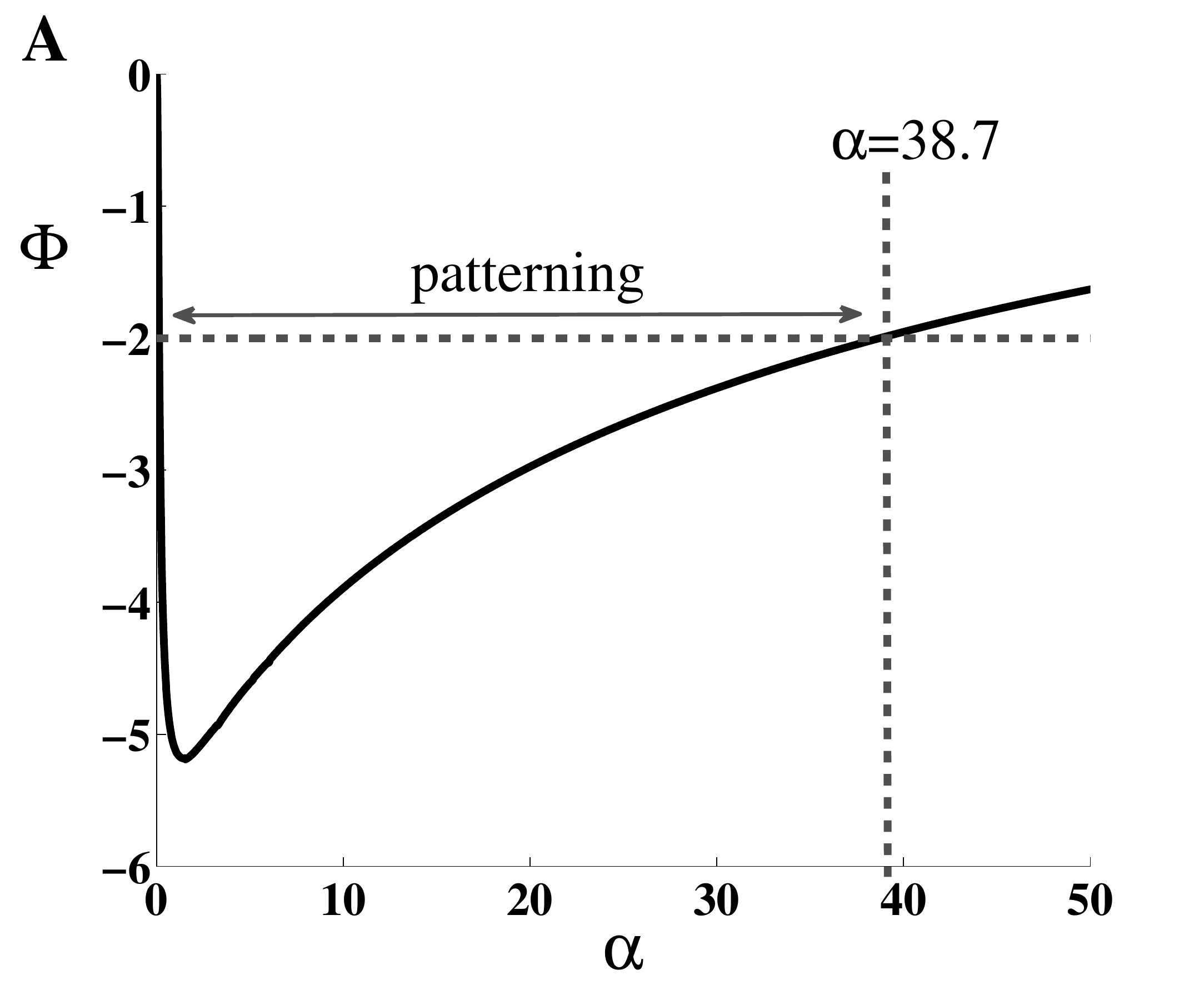} 
\label{fig:spr_morph_1}
\includegraphics[width=0.7\textwidth,angle=0]{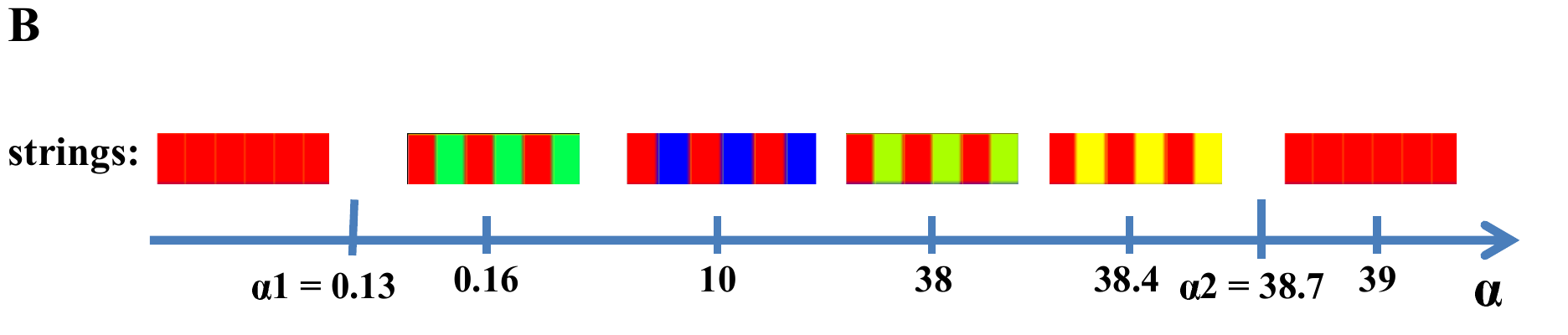} 
\label{fig:spr_morph_3}
\end{center}
\caption{\textbf{The range of parameter $\alpha$ for `salt-pepper' pattern formation.} (A) $\Phi$ as in equation (\ref{eq:FG_Collier}) over $\alpha$ (the maximum Delta production rate) showing the ranges of $\alpha$ for which we get the `salt-pepper' pattern.
(B) Simulation results for strings for different values of $\alpha$. Parameter values
used for our simulations as in Figure \ref{fig:eigenvalue}.
Colourbar: high (red) and low (blue) Delta levels at 1000 MCS.}\label{fig:determinant_over_alpha}
\end{figure}

\FloatBarrier
In Figure \ref{fig:determinant_over_alpha}(A) we plot $\Phi$ as a function of $\alpha$ (for fixed values of the other parameters), showing how $\Phi<-2$ for $\alpha \in [0.13,38.7]$, and hence patterning is predicted in that range.
%
Figure \ref{fig:determinant_over_alpha}(B) shows simulations as $\alpha$ is varied, giving patterning in the predicted range, and a stable homogeneous steady state otherwise.
%

\FloatBarrier

\FloatBarrier
\section*{Acknowledgments}
SAP has completed this work during his PhD at the University of Nottingham and a summer intern at Los Alamos National Laboratory, and acknowledges support from the Schools of Biosciences and Mathematical Sciences. This work supported by Award No. KUK-C1-013-04, made by King Abdullah University of Science and Technology (KAUST). YJ and LIA were partially supported by UC Lab Fees Research Grant: 09-LR-04-118348-IRU.


\FloatBarrier
\bibliographystyle{acm}
\bibliography{litteratureReview}

\end{document}